\documentclass[a4paper, 12pt]{article}

\usepackage{tikz}
\usepackage{latexsym,amsmath,amsfonts,amssymb}
\usepackage{amsthm}
\usepackage{mathtools}
\usepackage[latin1]{inputenc}
\usepackage[american]{babel}
\usepackage{bbm}
\usepackage[nosort]{cite}
\usepackage[colorlinks=true, citecolor=blue, linkcolor=blue, linktocpage=true]{hyperref}
\usepackage[symbol]{footmisc}

\renewcommand{\thefootnote}{\fnsymbol{footnote}}

\numberwithin{equation}{section}

\newcommand{\be}{\begin{equation}} \newcommand{\ee}{\end{equation}}
\newcommand{\bea}{\begin{equation} \begin{aligned}} \newcommand{\eea}{\end{aligned} \end{equation}}
\newcommand{\ba}{\begin{array}} \newcommand{\ea}{\end{array}}

\makeatletter
\def\blfootnote{\gdef\@thefnmark{}\@footnotetext}
\makeatother

\begin{document}

\thispagestyle{empty}

\vspace{12mm}
\begin{center}
{\large\bf Arm-Locking Frequency Noise Suppression for Astrodynamical Middle-Frequency Interferometric Gravitational Wave Observatory}
\\[15mm]
{Jun Nian$^{1}$ and Wei-Tou Ni$^{1,2}$}
 
\bigskip
{\it
$^1$ International Centre for Theoretical Physics Asia-Pacific,\\ University of Chinese Academy of Sciences, 100190 Beijing, China\\[.5em]
$^2$ Innovation Academy of Precision Measurement Science and Technology (APM),\\ Chinese Academy of Sciences,  430071 Wuhan,  China\\[.5em]
}

{\tt nianjun@ucas.ac.cn,  wei-tou.ni@wipm.ac.cn}

\bigskip
\bigskip

{\bf Abstract}\\[5mm]
{\parbox{14cm}{\hspace{5mm}
For space gravitational wave (GW) detection, arm locking is a proposal useful in decreasing the frequency noise of the laser sources for current developing space missions LISA and Taiji/TianQin.  In this paper, we study the application of arm locking to the Astrodynamical Middle-frequency Interferometric Gravitational wave Observatory (AMIGO) to decrease the frequency noise of laser sources. For AMIGO, the arm-locking technique can suppress the laser frequency noise by three orders of magnitude. The advantage of this is to make the auxiliary noise assignment for AMIGO easier and more relaxed. For the first-generation time-delay interferometry (TDI) configuration, the laser frequency noise contribution is already below the core noise contribution. For the simple Michelson TDI configuration (X0), the arm locking makes the acceleration-thrust scheme, the delay-line scheme, or the combined scheme easier to implement. Within a relatively short period of less than a day (compared to less than twenty days for LISA/Taiji), the Doppler frequency pulling can be efficiently reduced to within $\pm$ 0.001 Hz and does not affect the mission duty cycle much.}}

\end{center}

\newpage
\pagenumbering{arabic}
\setcounter{page}{1}
\setcounter{footnote}{0}
\renewcommand{\thefootnote}{\arabic{footnote}}

%{\renewcommand{\baselinestretch}{1} \parskip=0pt
%\setcounter{tocdepth}{2}
\tableofcontents

%%%%%%%%%%%%%%%%%%%%%%%%%%%%%%%%%%%%%%%%%%%%%%%%%%
%%%%%%%%%%%%%%%%%%%%%%%%%%%%%%%%%%%%%%%%%%%%%%%%%%

\section{Introduction}

\begin{figure}[!htb]
      \begin{center}
        \includegraphics[width=0.61\textwidth]{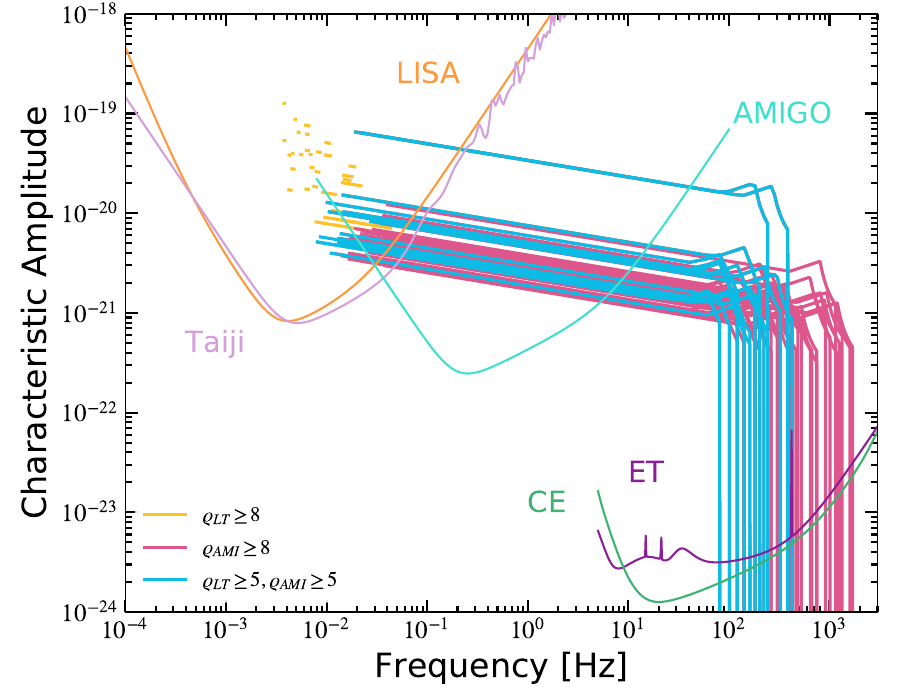}
        \caption{Evolution tracks of characteristic amplitudes for the mock BBHs which are detectable by LISA-Taiji-AMIGO (i.e.,  the signal-to-noise ratio for LISA-Taiji $\rho_{LT} \geq 5$ and for AMIGO $\rho_{AMI} \geq 5$,  blue lines), by LISA-Taiji (i.e.,  the signal-to-noise ratio for LISA-Taiji $\rho_{LT} \geq 8$,  yellow lines), and by AMIGO (i.e.,  the signal-to-noise ratio for AMIGO $\rho_{AMI} \geq 8$,  magenta lines), respectively. All the BBHs plotted here are observed for a continuous period of four years.  Orange, pink, sky blue, green, and purple curves represent the sensitivity curves for LISA, Taiji, AMIGO, CE, and ET, respectively \cite{Zhao:2023}.}
        \label{fig:BBHs}
      \end{center}
    \end{figure}

LIGO/Virgo/KAGRA (LVK) collaborations have reported the laser-interferometric detection of high-frequency ($\sim$ 10$\, - \,$1000 Hz) gravitational waves (GWs) emitted from more than 100 compact binary mergers,  i.e.,  stellar mass binary black holes (stellar BBH mergers),  binary neutron star (BNS) mergers,  and neutron star black hole (NSBH) mergers \cite{LIGOScientific:2021djp,  LIGOScientific:2016vlm,  LIGOScientific:2020iuh}.  In the low-frequency band,  LISA (Laser Interferometer Space Antenna for gravitational-wave detection) and Taiji (Taiji interferometer space antenna for gravitational-wave detection)/TianQin (TianQin interferometer space antenna for gravitational-wave detection) are actively developing and preparing laser-interferometric space missions for the GW sources of massive binary black holes, etc.  The mid-frequency band is between these two bands (0.1 Hz$\, - \,$10 Hz).  DECIGO (Deci-hertz Interferometer Gravitational wave Observatory) and a number of earth-based and other space-borne detectors are proposed for their GW sources (for a review of present status, see, e.g.,  \cite{Gao:2021esp}). According to a recent study \cite{Zhao:2023} (see also \cite{Cai:2023ywp}), the joint observation of LISA/Taiji, AMIGO (Astrodynamical Middle-frequency Interferometric Gravitational wave Observatory),  and CE (Cosmic Explorer)/ET (Einstein Telescope) with a connected binary black hole (BBH) inspiral template built from the evolution of massive binary stars in galactic field (EMBS channel) and the dynamical interactions in a dense stellar environment constrained by the GWTC-3 model \cite{KAGRA:2021duu} (Fig.~\ref{fig:BBHs}),  through mHz, deci-Hz, and deca-Hz frequencies, will be able to enhance the parameter estimation by two orders of magnitude or more (Tab.~\ref{tab:BBHs}).  These enhancements will strengthen the distinguishability of various GW source models, the precision of determination of cosmological models and Hubble constant, and the co-evolution of star formation, black holes, and galaxies. The sky localization will be unique for a significant number of sources to realize the corresponding redshift measurement. This will make a 0.1\% accuracy/precision cosmological model feasible to serve as a strong basis for cosmology and cosmological astrophysics. Each frequency band is crucial in reaching this important achievement.

\begin{table}[htb!]
\centering
\begin{tabular}{|c|c|c|c|c|}
\hline
GW detector & $\Delta \Omega_{90\%}$ & $\sigma_{d_L} / d_L$ & $\sigma_{M_c} / M_c$ & $\sigma_\eta$ \\
{} & Median & Median & Median & Median \\
\hline
LT & $8.2 \times 10^{-1}$ & $1.1 \times 10^{-1}$ & $3.4 \times 10^{-6}$ & $5.5 \times 10^{-3}$\\
\hline
AMIGO & $1.1 \times 10^{-1}$ & $7.5 \times 10^{-2}$ & $4.5 \times 10^{-7}$ & $4.7 \times 10^{-4}$ \\
\hline
ET-CE & $5.7 \times 10^{-3}$ & $1.8 \times 10^{-3}$ & $1.6 \times 10^{-3}$ & $2.0 \times 10^{-3}$ \\
\hline
LT-AMIGO & $5.4 \times 10^{-1}$ & $5.8 \times 10^{-2}$	 & $1.5 \times 10^{-7}$ & $3.0 \times 10^{-4}$ \\
\hline
LT-ET-CE & $1.5 \times 10^{-4}$ & $1.3 \times 10^{-3}$ & $4.7 \times 10^{-8}$ & $1.5 \times 10^{-4}$ \\
\hline
AMIGO-ET-CE & $4.9 \times 10^{-5}$ & $1.1 \times 10^{-3}$ & $2.4 \times 10^{-7}$ & $7.9 \times 10^{-5}$ \\
\hline
LT-AMIGO-ET-CE & $4.6 \times 10^{-5}$ & $1.1 \times 10^{-3}$ & $2.9 \times 10^{-8}$ & $6.1 \times 10^{-5}$ \\
\hline
\end{tabular}
\caption{Median values of the distributions of parameter estimation uncertainties for multiband BBHs \cite{Zhao:2023}.  The first column denotes the GW detector name or different combinations of them. The second column shows the median value of the distribution of sky localization $\Delta \Omega_{90\%}$ among 100 realizations of multiband BBHs. The third column, the fourth column, and the last column show the corresponding results for the distributions of relative uncertainty of luminosity distance $\sigma_{d_L} / d_L$,  the relative uncertainty of red-shifted chirp mass $\sigma_{M_c} / M_c$,  and the uncertainty of relative symmetric mass ratio $\sigma_\eta$,  respectively.  For the low and high values of the 68 percent confidence interval of each of these values,  see Ref.~\cite{Zhao:2023}.  \label{tab:BBHs}}
\end{table}

In the original white-light Michelson interferometry, the two interferometric paths (two arms) must be almost equal to have fringes \cite{Michelson1881relative}.  With the light of a very narrow line,  the match can be relaxed a bit (e.g.,  \cite{Michelson1925,  MichelsonGale1925}). For laser light, the interferometry goes further, and the fringes would manifest for a difference of optical paths within the laser coherent length. For Nd:YAG laser of 1 kHz linewidth, the two optical length paths can differ by 100 km and still be coherent to interfere with each other. However, the closer the path lengths are matched, the clearer the fringes are (less noise). In space, because of the large distances involved, at the receiving spacecraft (S/C), we have to phase lock the local laser oscillator to the weak incoming beam to transmit to another S/C or back; the interferometry measures the final phase of interference of the two chosen paths. The final phase noise $\delta \varphi_\text{interference}$ of interference at the receiving S/C is
\begin{align}\label{eq:delta varphi}
  \delta \varphi_\text{interference} & = 2 \pi \cdot \delta \nu (f)\cdot \Delta L/c + \text{phase locking noise(s)} + \text{timing noise(s)}\nonumber\\
  {} & \quad + \text{signals \& other noises accrued along the two paths}\, , 
\end{align}
where $\delta \nu (f)$ is the frequency noise of the laser source at frequency $f$,  $\Delta L$ is the optical path length difference of the two chosen paths, and $c$ is the light velocity.  Hence,  to decrease the interference phase noise $\delta \varphi_\text{interference}$,  we have to either decrease the frequency noise of the laser source or decrease the path length difference, or both. After the vast distance traveled, the light received by the telescope in the other S/C is attenuated greatly and needs amplification to go to another S/C. The amplification method uses a local laser to phase-lock to the incoming weak laser light. Hence, the phase information is transmitted in contiguous propagation, whether in homodyne or with a known frequency offset. The measured phase by the phasemeter is recorded with a time tag for later propagation identification in the data analysis. This time-tagged tracing after the recording is called time-delay interferometry (TDI). The first-generation Michelson TDI is usually called X, Y, and Z TDIs.  For X-TDI configuration, the two Michelson interference paths are: (i) S/C1 $\to$ S/C2 $\to$ S/C1 $\to$ S/C3 $\to$ S/C1; (ii) S/C1 $\to$ S/C3 $\to$ S/C1 $\to$ S/C2 $\to$ S/C1.  For Y-TDI,  start at S/C2; for Z-TDI, start at S/C3.  We call the original (zeroth generation) Michelson topology the X0, Y0, and Z0 TDIs.  That is for X0-TDI: (i) S/C1 $\to$ S/C2 $\to$ S/C1; (ii) S/C1 $\to$ S/C3 $\to$ S/C1; and so on.  The second-generation Michelson TDI is usually called X1,  Y1,  and Z1 TDIs.  For X1-TDI configuration,  the two Michelson interference paths are: (i) S/C1 $\to$ S/C2 $\to$ S/C1 $\to$ S/C3 $\to$ S/C1 $\to$ S/C3 $\to$ S/C1 $\to$ S/C2 $\to$ S/C1; (ii) S/C1 $\to$ S/C3 $\to$ S/C1 $\to$ S/C2 $\to$ S/C1 $\to$ S/C2 $\to$ S/C1 $\to$ S/C3 $\to$ S/C1; and so on.  For equal arms, two X0-TDI optical path lengths match each other,  and so on.  For constant arms,  two X-TDI optical path lengths match each other,  and so on.  For constant velocity formations,  two X1-TDI optical path lengths match,  and so on. Here, we summarize the definitions and some explanations of X, Y, Z, X1, Y1, Z1, X0, Y0, and Z0 TDIs. For further expositions of TDIs, please see \cite{Tinto:2020fcc, Wang:2020pkk, Wang:2020fwa} and the references therein. Both phase-locking and time-tagging in the intermediate S/C contribute to noises.  The TTL (tilt-to-length) noise,  etc.,  are included in the other noises.

For the present assumed laser source frequency noise of $\delta \nu (f) \leq 30\, \text{Hz} / \text{Hz}^{1/2}$,  the interference phase noise requirements of LISA and Taiji, the requirements of the path length difference $\Delta L$ are $\Delta L_\text{LISA} \leq 25\, \text{m}$ and $\Delta L_\text{Taiji} \leq 30\, \text{m}$,  respectively \cite{Wang:2017aqq}.  For the same laser source frequency noise requirement, the requirement of the path length difference is proportional to the arm length. For basically geodesic orbits of LISA, the variation of 2.5 Gm arm length can be 1\%,  i.e.,  25 Mm.  Hence,  for the simple (original) Michelson, the path length match can be one million times off the requirement.  For Taiji,  it can be 1 million times off, too. For two paths of TDI X configuration of LISA, the maximum path length difference can be less than $\pm 240\, \text{m}$ ($\pm 800\, \text{ns}$) for a mission time of 2200 days \cite{Wang:2017aqq}.  For two paths of TDI X configuration of Taiji,  the maximum path length difference can be less than $\pm 375\, \text{m}$ ($\pm 1250\, \text{ns}$) for a mission time of 2200 days. For two paths of the second generation TDI X1 configuration of LISA, the maximum path length difference can be less than $\pm 2.7\, \text{mm}$ ($\pm 9\, \text{ps}$) for 2200 days of mission time. For two paths of second generation TDI X1 configuration of LISA, the maximum path length difference can be less than $\pm 4.8\, \text{mm}$ ($\pm 16\, \text{ps}$) for 2200 days of mission time. For simple Michelson, i.e., the TDI X0 configuration and other first-generation TDIs of LISA and Taiji compiled in \cite{Wang:2017aqq},  the path length requirements are not satisfied. However, for the second-generation TDIs compiled in \cite{Wang:2017aqq},  the path length requirements are well satisfied.

Since the phase noise $\delta \varphi_\text{interference}$ of interference is also proportional to the frequency noise of the laser source, the other way to save the TDI X1 configuration for LISA and Taiji for GW detection is to decrease the frequency noise $\delta \nu (f)$ of the laser sources. This can be done by arm locking or using a more stable local laser system.

In this paper, we work out the arm locking schemes for the Astrodynamical Middle-frequency Interferometric Gravitational wave Observatory (AMIGO) \cite{Ni:2018,  Ni:2019nau,  Ni:2022}.  This paper is organized as follows.  Sec.~\ref{sec:AMIGO} gives a brief summary of AMIGO.  Sec.~\ref{sec:Noise} lists and discusses various noises assumed for this study.  Sec.~\ref{sec:SingleLocking} focuses on the single arm locking stabilization for AMIGO and AMIGO-5.  Sec.~\ref{sec:DoubleLocking} works out the double arm locking stabilization for AMIGO and AMIGO-5.  Sec.~\ref{sec:PDHLocking} presents the combined PDH cavity locking and double arm locking stabilization for AMIGO and AMIGO-5.  Sec.~\ref{sec:FrequencyPulling} considers the Doppler frequency pulling for this case.  Sec.~\ref{sec:Discussion} discusses various options in the context of mission implementation and what will be achieved in the network observations.

% \newpage
\section{Brief Review of AMIGO}\label{sec:AMIGO}

AMIGO (Astrodynamical Middle-frequency Interferometric Gravitational wave Observatory) is a laser-interferometric mission concept with three spacecraft forming a nearly equilateral triangular geodesic formation with nominal armlength of 10,000 km and $60^\circ$ nominal-inclination-to-ecliptical plane in Earth-like solar orbit interferometrically ranging with one another to detect mid-frequency (0.1-10 Hz) gravitational waves (GWs) from intermediate-mass binary black holes and EMRIs (extreme-mass-ratio-inspirals) to study formation channels of intermediate black holes, to study the evolution of globular clusters, to test relativistic gravity and to do multiband astronomy with low-frequency GW detectors and high-frequency detectors to explore precision cosmology \cite{Ni:2018,  Ni:2019nau,  Ni:2022}.  Tab.~\ref{tab:Parameters} lists key parameters of AMIGO and AMIGO-5 with 50,000 km nominal arm length.

\begin{table}[htb!]
\centering
\begin{tabular}{|c|c|c|}
\hline
Key parameters	 & AMIGO & AMIGO-5 \\
\hline
Arm length & 10,000 km & 50,000 km \\
\hline
Laser power & 2 W & 2 W \\
\hline
Laser wavelength & 1064 nm & 1064 nm \\
\hline
Telescope diameter & 0.3 m & 0.3 m \\
\hline
Arm laser metrology noise requirement & 3.8 fm Hz$^{-1/2}$ & 19 fm Hz$^{-1/2}$ \\
\hline
\end{tabular}
\caption{Key parameters of AMIGO and AMIGO-5 \cite{Ni:2022} \label{tab:Parameters}}
\end{table}

Since the noise requirement depends on development and budget, we have the b-AMIGO (basic AMIGO) version, which has relaxed arm laser metrology requirement, and e-AMIGO (enhanced AMIGO), which has more stringent arm laser metrology requirement \cite{Ni:2022}.

For the orbit configuration,  we use the AMIGO-S-8-12deg earth-like solar orbit trailing earth from 8 degrees to 12 degrees studied in \cite{Ni:2019nau}.  For the assumed laser source frequency noise of $\delta \nu (f) \leq 30\, \text{Hz} / \text{Hz}^{1/2}$ and the interference phase noise requirements of AMIGO, the requirement of the path length difference is $\Delta L_\text{AMIGO} \leq 0.1\, \text{m}$ \cite{Ni:2019nau}.  For the same laser source frequency noise requirement, the requirement of the path length difference is proportional to the arm length. For basically geodesic orbits of AMIGO, the arm length variation can be 1\%, i.e., 100 km, for 20 deg behind the earth or 20 deg leading the earth orbit configuration. For the AMIGO-S-8-12deg earth-like solar orbit trailing earth from 8 degrees to 12 degrees, the variation could be several times higher. Hence, for simple Michelson, the path length match can be several million times off. For two paths of the TDI X configuration of AMIGO, the maximum path length difference can be less than $\pm 6\, \text{mm}$ ($\pm 20\, \text{ps}$) for a mission time of 600 days. For the TDI X configuration and other non-Sagnac first-generation TDIs of AMIGO compiled in \cite{Ni:2019nau},  the path length requirement is satisfied (for the Sagnac TDI configurations, an extra path delay is needed for matching).

\begin{figure}[!htb]
      \begin{center}
        \includegraphics[width=0.78\textwidth]{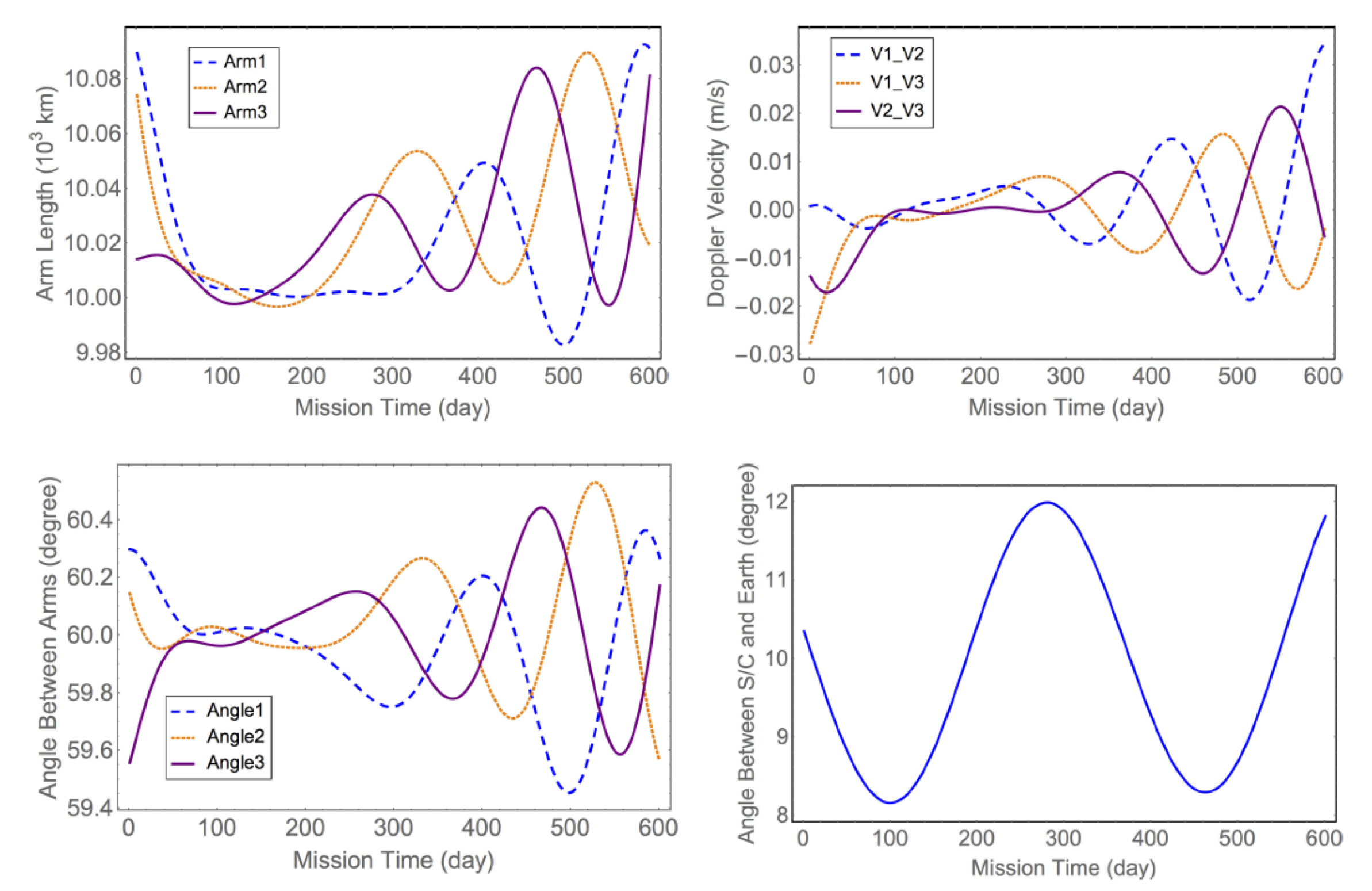}
        \caption{Variations of the arm lengths, the Doppler velocities, the formation angles,  and the angle between the barycenter of S/C and earth in 600 days for the AMIGO-S-8-12deg behind earth formation (from Ref.~\cite{Ni:2019nau}).}
        \label{fig:Orbits}
      \end{center}
    \end{figure}

In Ref.~\cite{Ni:2019nau},  the authors have studied the various orbit configurations, their Doppler velocities, and TDIs. Here, we use the AMIGO-S-8-12deg earth-like solar orbit formation with 8$\, - \,$12 degrees behind the earth for 600 days with initial conditions given in Column~3 of Tab.~1 of \cite{Ni:2019nau} as an example.  The arm length variations, the Doppler velocities, the formation angles, and the angle between the barycenter of S/C and earth in 600 days are shown in Fig.~\ref{fig:Orbits}.

\newpage
Doppler velocities for neighboring arms are less than 35 mm/s.  Path length differences of X0, Y0, Z0, X, Y, and Z TDIs and X+Y+Z are listed below in Tab.~\ref{tab:PathLength}.  As we can see, the requirement on $\Delta L$ for the first-generation TDIs is well satisfied by the orbit configurations (X,  Y,  Z,  and their sum).  According to Eq.~\eqref{eq:delta varphi},  the effective timing/tracking noise should also satisfy this requirement. If we set the effective timing/tracing noise to be below 100 ps, it would be OK. As to the phase-locking noise requirement, it is also well satisfied.  As we will see in this paper, the arm locking technique can efficiently suppress total noise by about three orders of magnitude in the mid-frequency range, making the requirement for the first-generation TDIs easier to satisfy. In the presence of the arm locking setup, the noise suppression performance is closer to the original Michelson topology (the zeroth-generation) TDIs. With the potential thrust-acceleration technology improvement, the arm locking technique will allow us to adopt only the Michelson topology TDIs, preventing additional complications.

\begin{table}[htb!]
\centering
\begin{tabular}{|c|c|c|c|}
\hline
TDI & Min & Max & Average \\
\hline
X0 (arm1 - arm2) & several hundred km & several hundred km & {}\\
\hline
Y0 (arm2 - arm3) & several hundred km & several hundred km & {}\\
\hline
Z0 (arm3 - arm1) & several hundred km & several hundred km & {}\\
\hline
X & $-19$ ps & $14$ ps & $7$ ps\\
\hline
Y & $-15$ ps & $14$ ps & $6$ ps\\
\hline
Z & $-12$ ps & $16$ ps & $6$ ps\\
\hline
X + Y + Z & $-0.01$ ps & $0.12$ ps & $0.06$ ps\\
\hline
Requirement on $\Delta L$ & $330$ ps & $330$ ps & {}\\
\hline
\end{tabular}
\caption{Minimum and maximum of path length differences for the Michelson configurations X0, Y0,  and Z0 TDIs, together with the minimum, maximum,  and average of the first-generation X, Y,  and Z TDI configurations plus those of the sum X+Y+Z are listed (from Ref.~\cite{Ni:2019nau}).  \label{tab:PathLength}}
\end{table}

% \newpage
\section{Noise Characterization}\label{sec:Noise}

In this section,  we set the noise requirements for AMIGO and list the noise models considered in this paper.

For the acceleration noise requirement, we set
\be\label{eq:AccNoiseRequirementAMIGO}
  S_a^{1/2} (f) \leq 3 \times 10^{-15} \left[ 1 + \left(\frac{f}{0.3\, \text{Hz}} \right)^4 \right]^{1/2} \frac{\text{m}}{\text{s}^2\, \text{Hz}^{1/2}}\, ,\quad (10\, \text{mHz} < f < 10\, \text{Hz})\, ,
\ee
for b-AMIGO, AMIGO and e-AMIGO.  N.B.,  LISA Pathfinder (LPF) has already demonstrated the following requirement in the frequency range $20\, \mu\text{Hz} - 0.03\, \text{Hz}$ in Feb 2017, 
\be\label{eq:AccNoiseRequirementLPF}
  S_a (f) \leq 9 \times 10^{-30} \left[1 + \left(\frac{10^{-4}\, \text{Hz}}{f} \right)^2 + 16\, \left(\frac{2 \times 10^{-5}\, \text{Hz}}{f} \right)^{10} \right] \frac{\text{m}^2}{\text{s}^4\, \text{Hz}}\, , \quad (20\, \mu\text{Hz} - 0.03\, \text{Hz})\, .
\ee
With the achieved requirement \eqref{eq:AccNoiseRequirementLPF} of LISA LPF,  our requirement \eqref{eq:AccNoiseRequirementAMIGO} is already satisfied in $10\, \text{mHz}$ to $30\, \text{mHz}$ frequency band.  For the $0.03\, \text{Hz}$ to $10\, \text{Hz}$ band,  the flat requirement $S_a^{1/2} (f) \leq 3 \times 10^{-15}\, \text{m}\, \text{s}^{-2}\, \text{Hz}^{-1/2}$ should be no more difficult to implement.  However, we relax this by a blue factor with a corner frequency $[1 + ( f / 0.3\, \text{Hz})^4]^{1/2}$.  Since there is already an antenna factor in sensitivity,  the blue factor will not affect the sensitivity significantly.

For the laser metrology noise, we set
\begin{itemize}
\item Baseline (b-AMIGO):
\be
  S_\text{AMIGOp} \leq 1.4 \times 10^{-28}\, \frac{\text{m}^2}{\text{Hz}}\, ,\quad S_\text{AMIGOp}^{1/2} \leq 12\, \frac{\text{fm}}{\text{Hz}^{1/2}}\, , \quad (10\, \text{mHz} < f < 10\, \text{Hz})\, ;
\ee
                        
\item Design Goal (AMIGO):
\be\label{eq:AMIGOshot}
  S_\text{AMIGOp} \leq 0.14 \times 10^{-28}\, \frac{\text{m}^2}{\text{Hz}}\, ,\quad S_\text{AMIGOp}^{1/2} \leq 3.8\, \frac{\text{fm}}{\text{Hz}^{1/2}}\, ,\quad (10\, \text{mHz} < f < 10\, \text{Hz})\, ;
\ee

\item Enhanced Goal (e-AMIGO):
\be
  S_\text{AMIGOp} \leq 0.0025 \times 10^{-28}\, \frac{\text{m}^2}{\text{Hz}}\, ,\quad S_\text{AMIGOp}^{1/2} \leq 0.5\, \frac{\text{fm}}{\text{Hz}^{1/2}}\, ,\quad (10\, \text{mHz} < f < 10\, \text{Hz})\, .
\ee
\end{itemize}
As to laser frequency noise, we use the same criterion as LISA/Taiji,  but in the  $10\, \text{mHz} < f < 10\, \text{Hz}$ band:
\be\label{eq:LaserAfterPreStabilization}
  \delta \nu (f) \leq 30\, \frac{\text{Hz}}{\text{Hz}^{1/2}}\, ,\quad (10\, \text{mHz} < f < 10\, \text{Hz})\, . 
\ee
This is the residual laser frequency noise after the pre-stabilization,  e.g.,  through the PDH cavity discussed below.  As a comparison,  the stabilized laser system of the GRACE follow-on has the following laser frequency noise \cite{Bachman:2017,  Ghosh:2021adg}: 
\begin{align}
  \delta \nu (f) & \approx 0.4\, \left(\frac{1\, \text{Hz}}{f}\right) \left[1 + \left(\frac{1\, \text{mHz}}{f}\right)^2\right]^{3/4} \frac{\text{Hz}}{\text{Hz}^{1/2}} \nonumber\\
  {} & \approx 0.4\, \left(\frac{1\, \text{Hz}}{f}\right) \frac{\text{Hz}}{\text{Hz}^{1/2}}\, .
\end{align}

More explicitly,  we list a few noises considered in this paper.
\begin{itemize}
\item Laser frequency noise:

The laser frequency noise for both AMIGO and AMIGO-5 is given by the same model as LISA \cite{McKenzie:2009kt,  Valliyakalayil:2021jxd}:
\be
  \nu_L\left(f\right)=\frac{30000\ \text{Hz}}{f}\frac{\text{Hz}}{\sqrt{\text{Hz}}}\ .
\ee
This expression characterizes the typical laser frequency noise in a free-running non-planar ring oscillator laser.  In contrast,  \eqref{eq:LaserAfterPreStabilization} is for the laser frequency noise after pre-stabilization.

\item Shot noise:

We can compare the design goal for the laser metrology noise of AMIGO given by \eqref{eq:AMIGOshot} to the value of LISA,  $10\, \text{pm} / \text{Hz}^{1/2}$.  The shot noise for AMIGO can then be obtained by rescaling the shot noise model of LISA \cite{McKenzie:2009kt,  Valliyakalayil:2021jxd}:
\be
  \nu_{shot} (f) = (3.8 \times 10^{-4}) \cdot (6.9 \times 10^{-6})\cdot \frac{2\pi i f}{1\, \text{Hz}}\, \frac{\text{Hz}}{\sqrt{\text{Hz}}}\, .
\ee
Similarly,  the shot noise for AMIGO-5 is
\be
  \nu_{shot} (f) = (1.9 \times 10^{-3}) \cdot (6.9 \times 10^{-6})\cdot \frac{2\pi i f}{1\, \text{Hz}}\, \frac{\text{Hz}}{\sqrt{\text{Hz}}}\, .
\ee

\item Clock noise:

We take the clock noise for both AMIGO and AMIGO-5 to be the same as LISA \cite{McKenzie:2009kt,  Valliyakalayil:2021jxd}:
\be
  \nu_{clock}\left(f\right)=(30\ \text{MHz})\cdot C_i\left(f\right)\cdot \frac{2\pi i f}{1\, \text{Hz}}\, \frac{\text{Hz}}{\sqrt{\text{Hz}}}\ ,
\ee
where $C_i\left(f\right)$ is given by
\be
  C_i\left(f\right)=\frac{2.4\times{10}^{-12}}{2\ \pi\ \left(f / (1\, \text{Hz}) \right)^{3/2}}\, \frac{1}{\text{Hz}}\ .
\ee

\item Spacecraft motion noise:

The spacecraft motion noise for both AMIGO and AMIGO-5 has the same form as LISA \cite{PhysRevD.90.062005,  Valliyakalayil:2021jxd}:
\be
  \nu_{SC}\left(f\right)=\phi_{SC}\left(f\right)\cdot 2\pi i\cdot f\, \frac{\text{Hz}}{\sqrt{\text{Hz}}}\ ,
\ee
where $\phi_{SC}\left(f\right)$ is given by
\be
  \phi_{SC}\left(f\right)=\frac{1.5\ \sqrt{1+\left(8\ \text{mHz} / f\right)^4}\ \text{nm}}{1064\ \text{nm}}\ \frac{1}{\text{Hz}}\ .
\ee

\item PDH cavity noise:

The Pound-Drever-Hall (PDH) cavity can be used in the laser pre-stabilization,  but it also introduces noise for both AMIGO and AMIGO-5 is given by the same form as LISA \cite{Valliyakalayil:2021jxd}:
\be
  \nu_{cavity}\left(f\right)=30\ \sqrt{1+\left(\frac{2\ \text{mHz}}{f}\right)^4}\frac{\text{Hz}}{\sqrt{\text{Hz}}}\ .
\ee

\item The first-generation TDI's requirement:

Because we eventually compare the noise suppression after the arm locking with the first-generation TDI's requirement,  we also list the first-generation TDI's requirement,  which is obtained by extrapolating the model for LISA \cite{Valliyakalayil:2021jxd} to the mid-frequency range (e.g.,  AMIGO):
\be
  \nu_{TDI-1} (f) = 425\, \sqrt{1+\left(\frac{2\ \text{mHz}}{f}\right)^4}\, \frac{\text{Hz}}{\sqrt{\text{Hz}}}\, .
\ee

\end{itemize}

\section{Single Arm Locking}\label{sec:SingleLocking}

In space-based gravitational wave detection,  several spacecrafts form an interferometer,  and the distances between each two of them are the arms of the interferometer.  Compared to the laser frequency noise,  the arm lengths are relatively more stable in this case.  Following the idea in \cite{Sheard:2003} (see also \cite{PhysRevD.90.062005,  Ghosh:2021adg}),  we can use the stability of the arm length to suppress the laser frequency noise.  This technique is called arm locking,  and single arm locking denotes the model where only one arm is considered.

The basic idea of arm locking is constructing a feedback control system by sending the amplified signal back into the original laser after it travels around the arm once. The interference responds to the difference between the phases of the beam returning from the distant spacecraft and the local oscillator derived from the outgoing laser \cite{Shaddock_2008}. The phase fluctuations of the local laser are sensed with no delay, making a high bandwidth system possible, as first noted in Ref.~\cite{Sheard:2003} (see also Ref.~\cite{2006PhD} and the references therein).

As a consequence,  the noises in the system are not given by the original bare noise models.  Instead,  they should be solved within the closed-loop feedback system.  A concrete analytical calculation shows that the new laser frequency noise can be efficiently suppressed within this closed-loop feedback system compared to a free-running laser.  At the same time, the gravitational wave signal maintains its magnitude.

\begin{figure}[!htb]
      \begin{center}
        \includegraphics[width=0.95\textwidth]{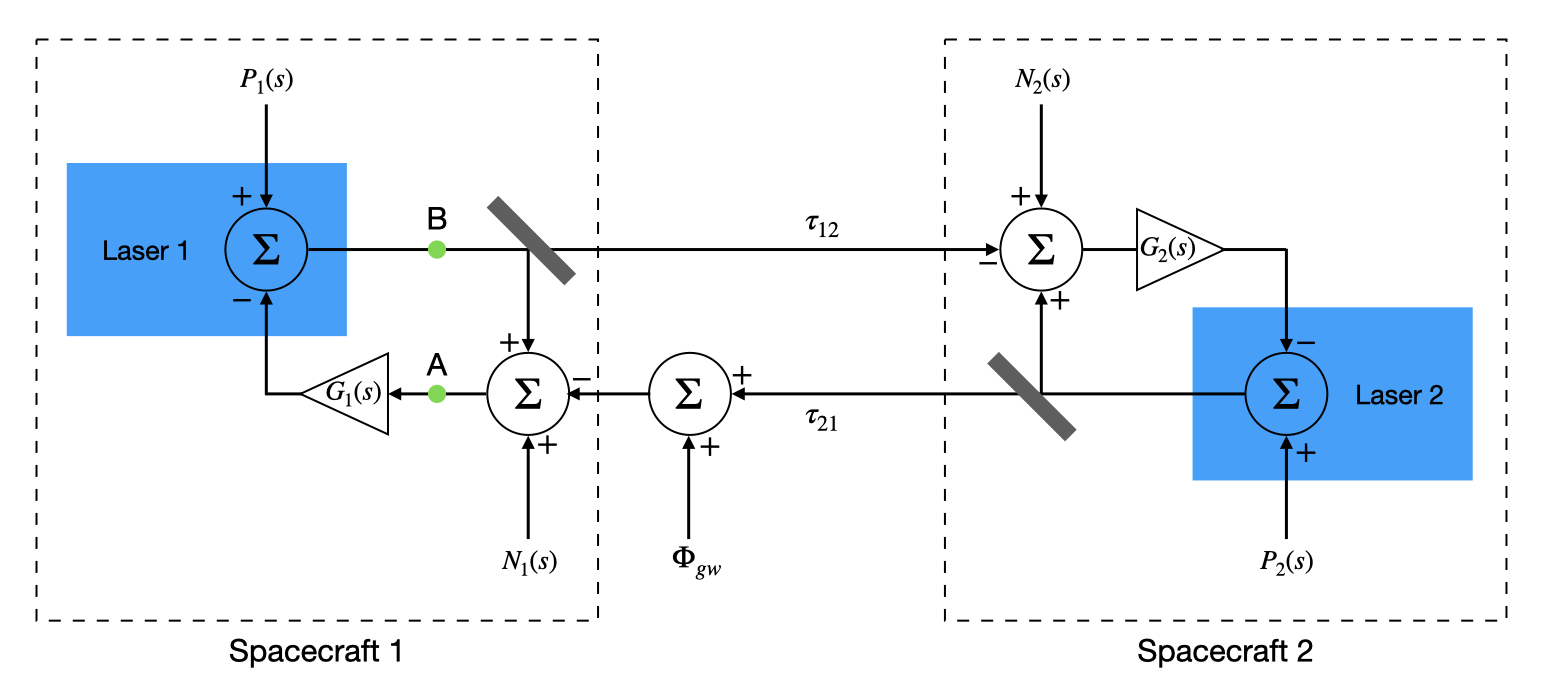}
        \caption{The feedback system of the single arm locking (see also \cite{Sheard:2003})}
        \label{fig:SingleArmLockingStructure}
      \end{center}
    \end{figure}

In this section,  we consider the single arm locking with the feedback system shown in Fig.~\ref{fig:SingleArmLockingStructure} and discuss how the noises mentioned in the previous section are suppressed after the arm locking.  For the single arm locking,  the close-loop laser noise at Point $B$ after the arm locking is given by \cite{Sheard:2003}:
\be\label{eq:SingleArmLocking}
  \mathrm{\Phi}_{\text{single}}=\frac{P_1}{1+L_1}-\frac{G_1\ P_2}{\left(1+G_2\right)\left(1+L_1\right)}-\frac{G_1G_2N_2}{\left(1+G_2\right)\left(1+L_1\right)}-\frac{G_1\ N_1}{1+L_1}+\frac{G_1\ \mathrm{\Phi}_{gw}}{1+L_1} ,
\ee
where $L_1=\left(1-G_2\ e^{-s\, \tau}/(1+G_2)\right)G_1$ with $L_1$ denoting the open-loop transfer function. In this expression,  $P_{1,  2}$ and $N_{1,  2}$ are the laser frequency noises and the shot noises on Spacecraft 1 and 2, respectively,  while $\mathrm{\Phi}_{gw}$ denotes the gravitational wave signal. We can tune the arm locking controllers $G_{1,  2}$ to adjust the final noise $\mathrm{\Phi}_{\text{single}}$ after the single arm locking.

In the large-gain limit, i.e., very large $G_{1, 2}$, we see that $L_1 \to G_1$. Hence, in the large-gain limit, the first two terms in \eqref{eq:SingleArmLocking} standing for the laser frequency noises are suppressed,  while the other terms remain the same order. In particular,  the gravitational wave signal $\Phi_{gw}$ is not suppressed in this limit, in contrast to the laser frequency noises $P_1$ and $P_2$ which are suppressed by the large gains $G_{1,2}$ as $L_1 \to G_1$. However, a single-arm locking setup is insufficient to extract gravitational wave signals, for which multi-arm configurations are indispensable. When two arms, each with a single-arm-locked laser, are available, the gravitational wave signal can be obtained by comparing the single-arm-locked lasers for different arms using a phase meter at point B, which is different from a double-arm locking configuration discussed in the next section.

Depending on the choices of the controllers $G_{1, 2}$,  there are two classes of the single arm locking: (i) the symmetric controller solution when $G_1=G_2$; (ii) the asymmetric controller solution when $G_1\neq G_2$.

Here,  we adopt and modify the controllers designed for LISA \cite{Sheard:2003}.  The values of the parameters are fixed by optimizing the laser frequency noise suppression in the mid-frequency range (0.1 Hz$\, - \,$10 Hz), which are shown in Tab.~\ref{tab:SingleLockingController}. The controllers $G_{1, 2}$ in Tab.~\ref{tab:SingleLockingController} contain a factor $1/\sqrt{f}$, which cannot be easily implemented in time domain simulations. The same problem appears in the previous literature \cite{McKenzie:2009kt,  PhysRevD.90.062005,  Valliyakalayil:2021jxd}, and the resolution is to use multiple low-pass filters with appropriate zeros, poles, and gains to approximate $1/\sqrt{f}$ in an expected frequency range for the time-domain simulations. In this paper, we focus on the frequency domain, so we keep the factor $1/\sqrt{f}$ in the computations for simplicity.

\begin{figure}[!htb]
      \begin{center}
        \includegraphics[width=0.75\textwidth]{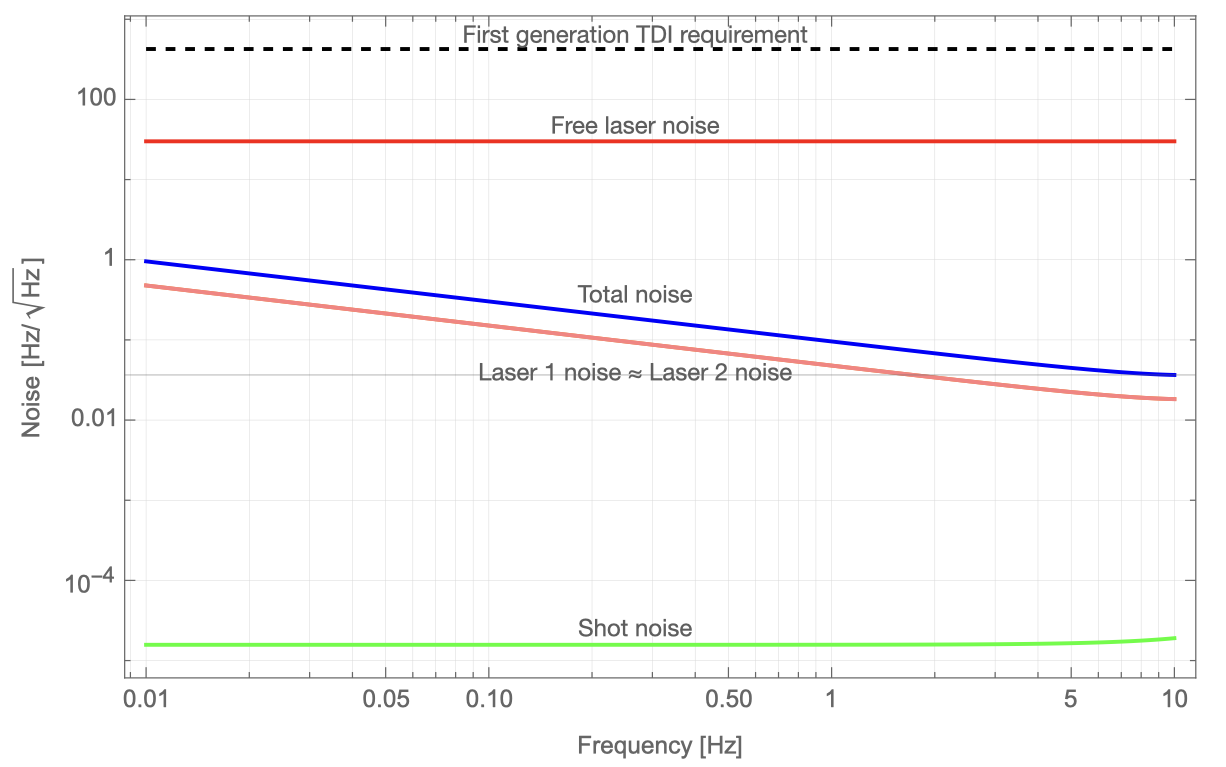}
        \caption{The results of the single arm locking with symmetric controllers for AMIGO}
        \label{fig:SingleLockingAMIGOSymm}
      \end{center}
    \end{figure}

\begin{figure}[!htb]
      \begin{center}
        \includegraphics[width=0.75\textwidth]{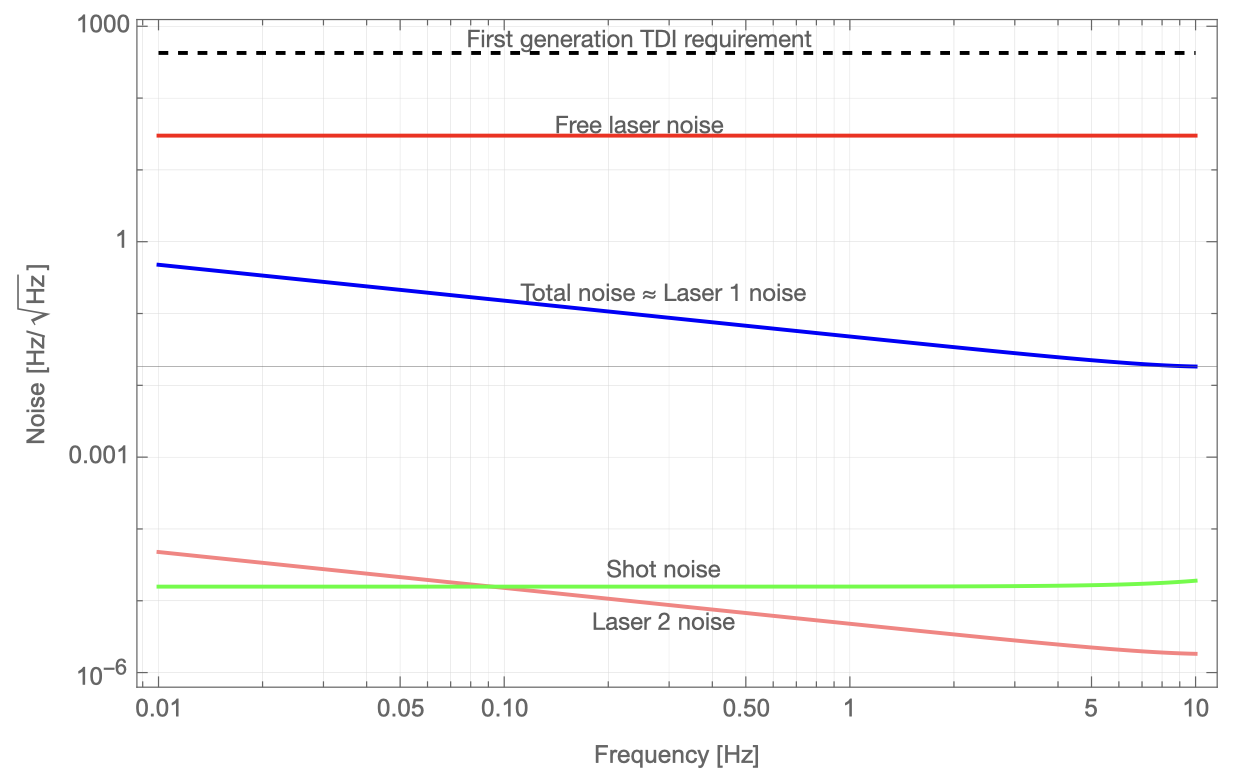}
        \caption{The results of the single arm locking with asymmetric controllers for AMIGO}
        \label{fig:SingleLockingAMIGOAsymm}
      \end{center}
    \end{figure}

\begin{figure}[!htb]
      \begin{center}
        \includegraphics[width=0.75\textwidth]{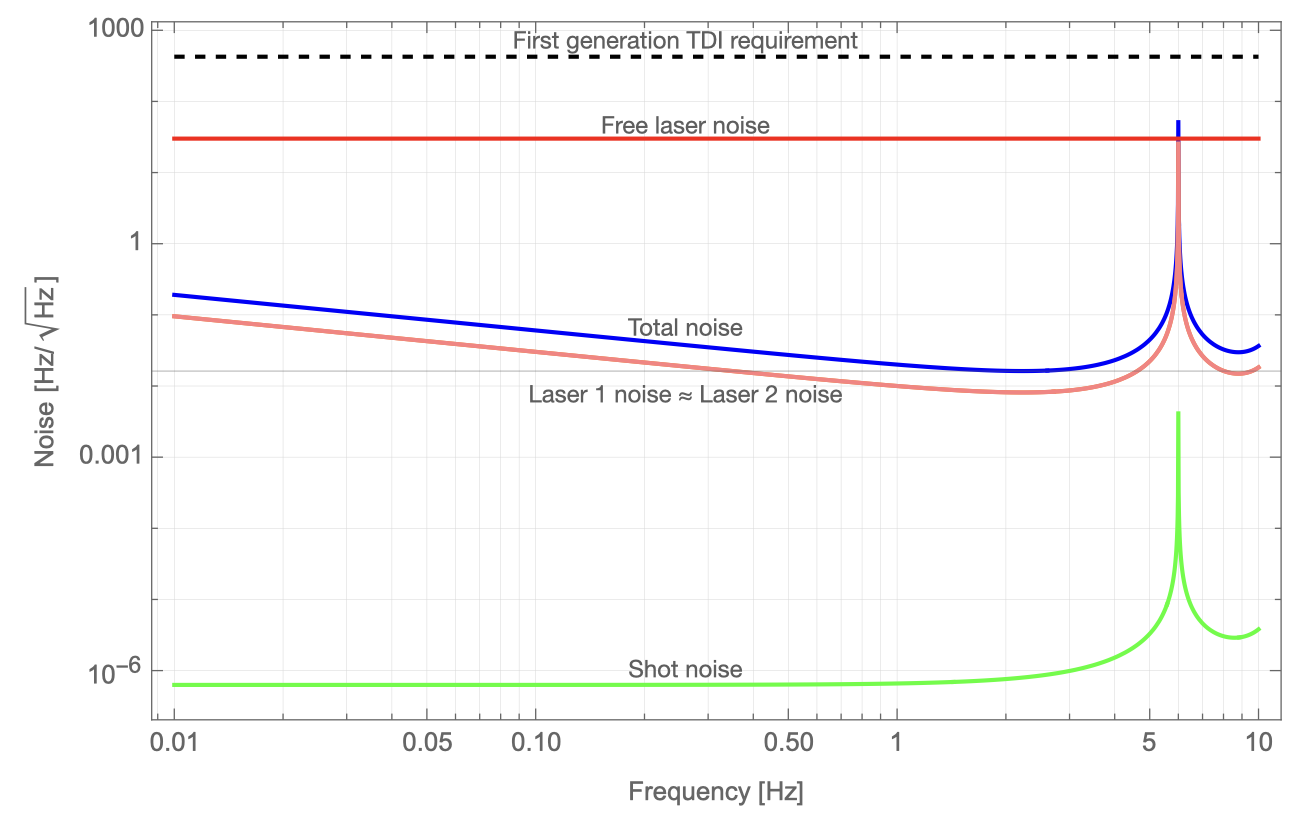}
        \caption{The results of the single arm locking with symmetric controllers for AMIGO-5}
        \label{fig:SingleLockingAMIGO-5Symm}
      \end{center}
    \end{figure}

\begin{figure}[!htb]
      \begin{center}
        \includegraphics[width=0.75\textwidth]{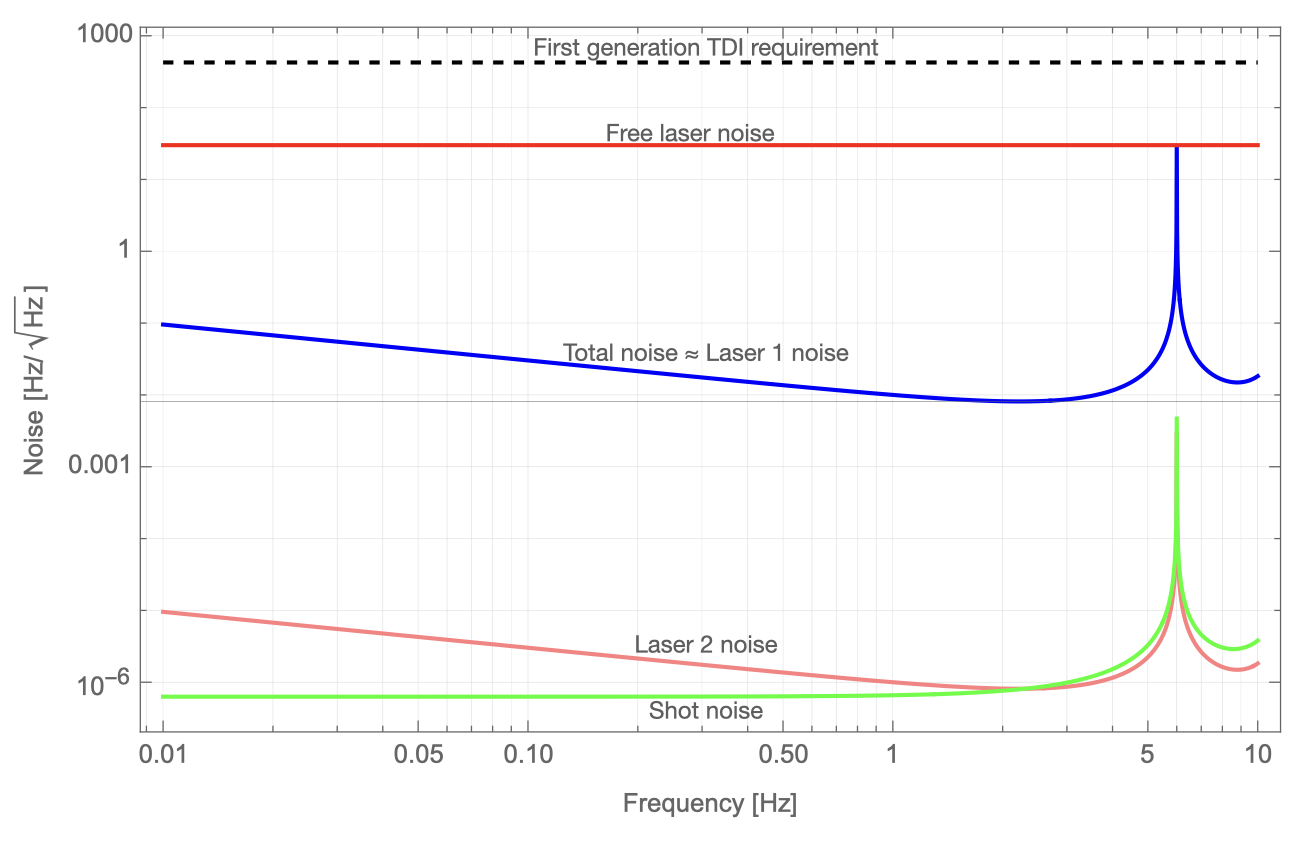}
        \caption{The results of the single arm locking with asymmetric controllers for AMIGO-5}
        \label{fig:SingleLockingAMIGO-5Asymm}
      \end{center}
    \end{figure}

After turning on the single arm locking, the noises for AMIGO (arm length 10,000 km) and AMIGO-5 (arm length 50,000 km) are shown in Fig.~\ref{fig:SingleLockingAMIGOSymm} $-$ Fig.~\ref{fig:SingleLockingAMIGO-5Asymm}. In these figures, the free laser noise (after pre-stabilization) and the requirement of the first-generation TDI are denoted by the red line and the black dotted line,  respectively.  The orange, green, and blue lines indicate the laser frequency noise after the single arm locking, the shot noise, and the total noise, respectively.

\begin{table}[htb!]
\centering
\begin{tabular}{|c|c|c|}
\hline
{} & $G_1 [f]$ & $G_2 [f]$ \\
\hline
Symmetric & $10^5/ \sqrt{f\, [\text{Hz}]}$ & $10^5/ \sqrt{f\, [\text{Hz}]}$\\
\hline
Asymmetric & $10^5/ \sqrt{f\, [\text{Hz}]}$ & $10^8/ \sqrt{f\, [\text{Hz}]}$\\
\hline
\end{tabular}
\caption{The choices of the controllers $G_{1, 2}$ for the single arm locking \label{tab:SingleLockingController}}
\end{table}

\newpage
\
\newpage
After the single arm locking, we see that the laser frequency noise can be suppressed for two to four orders of magnitude in the mid-frequency range (10 mHz$\, - \,$10 Hz).

\section{Double Arm Locking}\label{sec:DoubleLocking}

\begin{figure}[!htb]
      \begin{center}
        \includegraphics[width=0.95\textwidth]{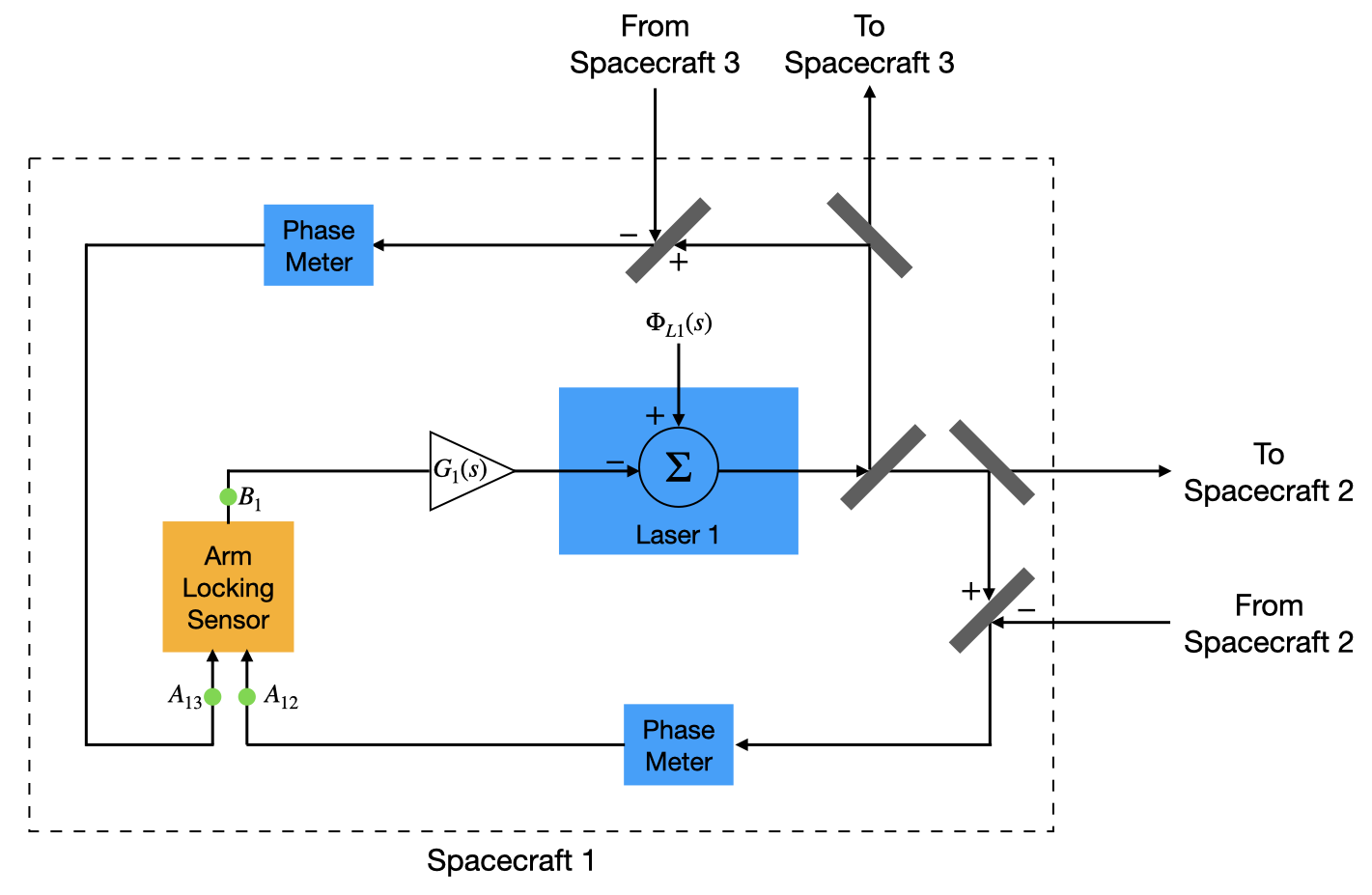}
        \caption{The feedback system of the double arm locking (see also \cite{McKenzie:2009kt})}
        \label{fig:DoubleArmLockingStructure}
      \end{center}
    \end{figure}

In gravitational wave detection, there are multiple arms in the configuration. We can combine the signals from different arms before feeding them back to the original laser. This can be done by an arm-locking sensor, which can be viewed as a linear operation of the signals from two arms (see Fig.~\ref{fig:DoubleArmLockingStructure}):
\be
  \phi_{double}\left(\omega\right)= \mathbf{S} \left(\begin{matrix}\phi_{13}\left(\omega\right)\\\phi_{12}\left(\omega\right)\\\end{matrix}\right) .
\ee
When two arms participate in the feedback control system,  this way of arm locking is called double arm locking \cite{McKenzie:2009kt} (see also \cite{PhysRevD.90.062005,  Ghosh:2021adg}).  Depending on the linear combinations of the signals from the two arms, the double arm locking includes several types (common, dual, modified dual, etc.).
   \\
For simplicity, in this paper, we only consider the common double arm locking, which is just taking the sum of the signals from two arms, i.e.,  $\mathbf{S}_+=(1,\ \ 1)$.  Using the common double arm locking, the closed-loop laser noise after the arm locking is given by \cite{McKenzie:2009kt}:
\be
  \nu_O=\frac{\nu_L}{1+G_1\ P_+\ }-\frac{G_1\ \mathbf{S}_+\ {(N}_{SN}+N_{CN}+N_{SCN})}{1+G_1\ P_+\ }\ ,
\ee
where $\nu_L$ is the laser frequency noise,  $N_{SN}$,  $N_{CN}$,  $N_{SCN}$ denote the shot noise, the clock noise, and the spacecraft motion noise (the optical bench displacement noise),  respectively.  The choices of the controllers $G_{1,2}$ and the frequency response $P_+$ are given in Tabs.~\ref{tab:DoubleLockingController-1} and~\ref{tab:DoubleLockingController-2}.

\begin{table}[htb!]
\centering
\begin{tabular}{|c|c|}
\hline
$G_1 [s]$ & $\left(\frac{g_1}{s}\right)^{2.3}\ \left(\frac{s}{s+p_{h_1}}\right)^5\ \left(\frac{s}{s+p_{h_2}}\right)^2\ \left(g_{l_c}\left(\frac{s+z_{l_c}}{s+p_{l_c}}\right)\right)$ \\
\hline
$G_2 [s]$ & $\left(g_2 / s\right)^{1.5}$ \\
\hline
$P_+ [\omega]$ & $2\ \left[1-\cos{\left(\Delta\tau\ \omega\right)}\ e^{-i\ \omega\ \bar{\tau}}\right]\qquad
(\bar{\tau}\equiv\tau_{12}+\tau_{13},\ \ \Delta\tau\equiv\tau_{12}-\tau_{13})$ \\
\hline
\end{tabular}
\caption{The controllers for the common double arm locking \label{tab:DoubleLockingController-1}}
\end{table}

\begin{table}[htb!]
\centering
\begin{tabular}{|c|c|}
\hline
Parameter & Value\\
\hline
$g_1$ & $2\pi\times0.68\times{10}^5$ \\
\hline
$g_2$ & $2\pi\times7.32\times{10}^3$ \\
\hline
$p_{h_1}$ & $2\pi\times3.87\times{10}^{-5}$ rad/s \\
\hline
$p_{h_2}$ & $2\pi\times1.16\times{10}^{-2} $ rad/s \\
\hline
$z_{l_c}$ & $2\pi\times{10}^{-5}$ rad/s \\
\hline
$p_{l_c}$ & $2\pi\times1.35\times{10}^{-4}$ rad/s \\
\hline
$g_{l_c}$ & $0.045$\\
\hline
\end{tabular}
\caption{The parameters for the common double arm locking \label{tab:DoubleLockingController-2}}
\end{table}

\begin{figure}[!htb]
      \begin{center}
        \includegraphics[width=0.75\textwidth]{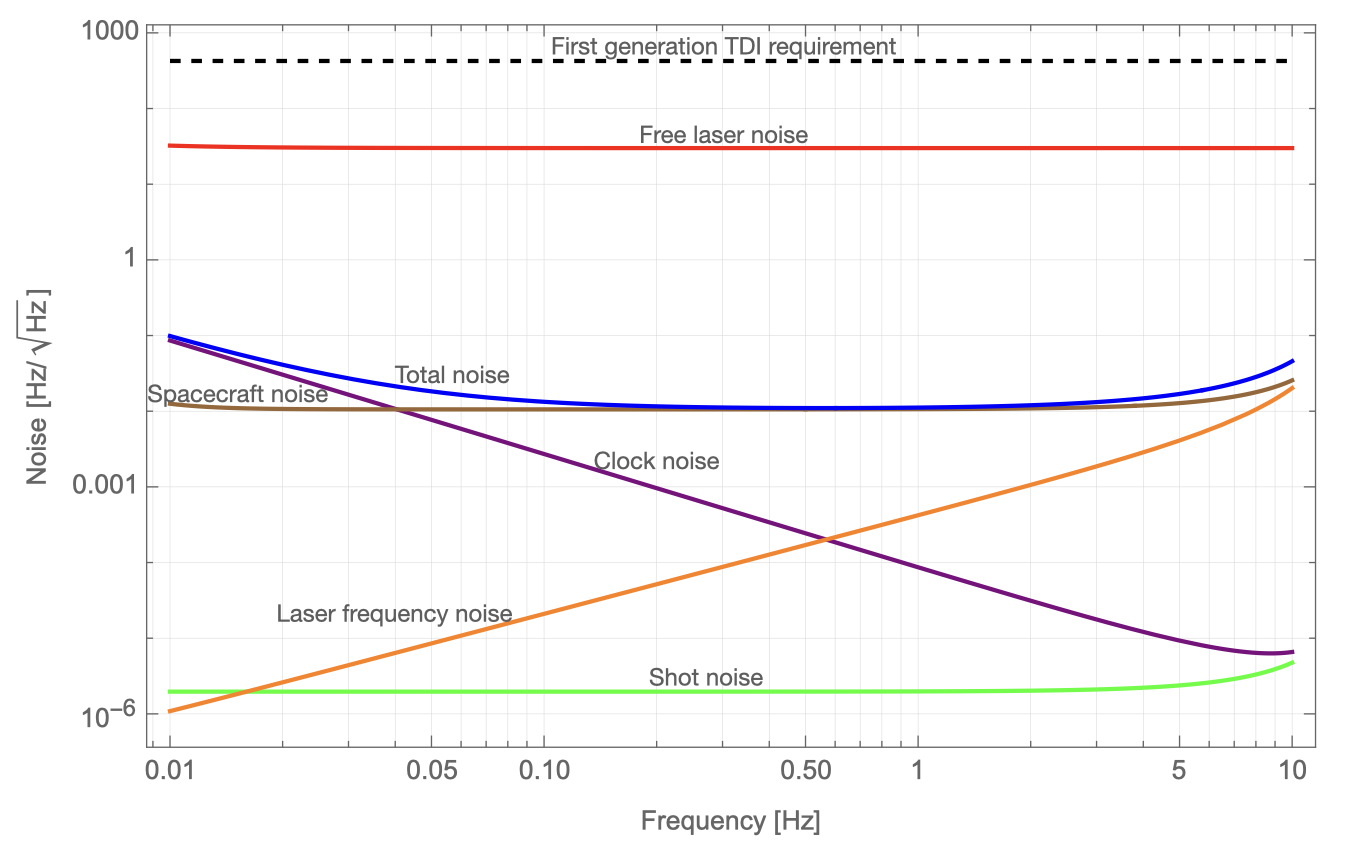}
        \caption{The noise budget of the common double arm locking for AMIGO}
        \label{fig:DoubleLockingAMIGO}
      \end{center}
    \end{figure}

\begin{figure}[!htb]
      \begin{center}
        \includegraphics[width=0.75\textwidth]{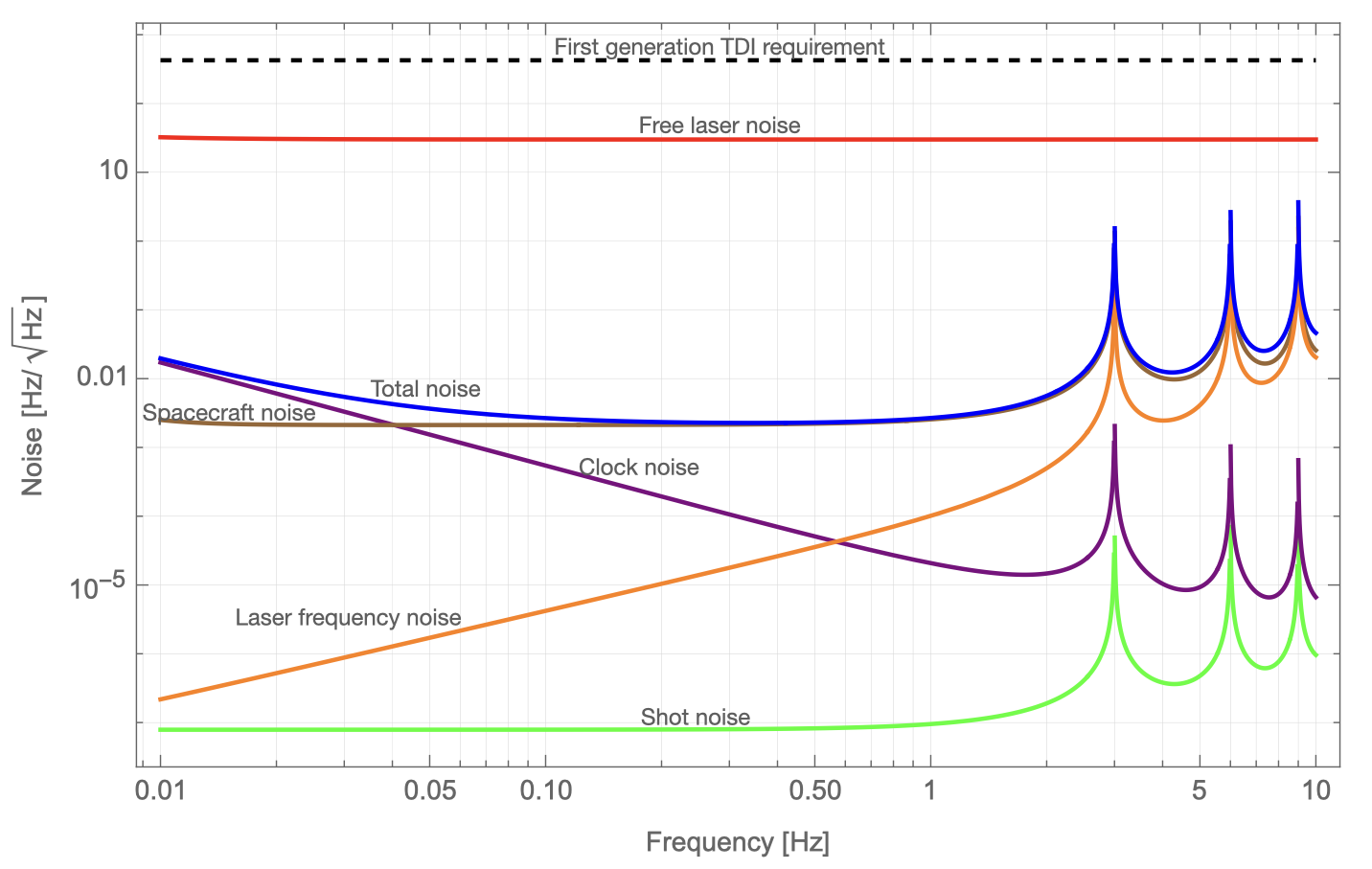}
        \caption{The noise budget of the common double arm locking for AMIGO-5}
        \label{fig:DoubleLockingAMIGO-5}
      \end{center}
    \end{figure}

The results after the double arm locking are shown in Figs.~\ref{fig:DoubleLockingAMIGO} and \ref{fig:DoubleLockingAMIGO-5}. Besides the free laser noise (red line) and the first-generation TDI requirement (black dotted line), we also include the laser frequency noise after the double arm locking (orange line),  the shot noise (green line),  the clock noise (violet line),  the spacecraft motion noise (brown line),  and the total noise (blue line).  We assume the average arm lengths for AMIGO and AMIGO-5 are 10,000 km and 50,000 km,  respectively, and the difference between the two arms can be up to 1\%.

\newpage

With the common double arm locking,  the noise budgets for AMIGO and AMIGO-5 are plotted in Fig.~\ref{fig:DoubleLockingAMIGO} and \ref{fig:DoubleLockingAMIGO-5}.  As we can see from these figures,  the total noise can be suppressed to meet the requirement of the first-generation TDI in the mid-frequency range.

\newpage
\section{Combined PDH Cavity and Double Arm Locking}\label{sec:PDHLocking}

On top of the double arm locking, we can introduce a Pound-Drever-Hall (PDH) cavity to enhance the laser frequency noise suppression.  In the presence of the PDH cavity, the closed-loop laser noise after the arm locking is given by \cite{Valliyakalayil:2021jxd}:
\be
  \nu_{PDH}=\frac{\nu_L}{1+G_1\ P_++G_2\ P_{pdh}}-\frac{G_2\ P_{pdh}\ \nu_{cavity}}{1+G_1\ P_++G_2\ P_{pdh}}-\frac{G_1\ \mathbf{S}_+\ {(N}_{SN}+N_{CN}+N_{SCN})}{1+G_1\ P_++G_2\ P_{pdh}}\, ,
\ee
where the transfer function of the PDH cavity is given by
\be
  P_{pdh\ }=\frac{D_0}{1+\frac{s}{2\ \pi\ f_c}}
\ee
with a dimensionless constant $D_0$ and $f_c\approx100\ kHz$.  We use the same controllers $G_{1, 2}$ with the same parameters as in the common double arm locking discussed in the previous section.

\begin{figure}[!htb]
      \begin{center}
        \includegraphics[width=0.75\textwidth]{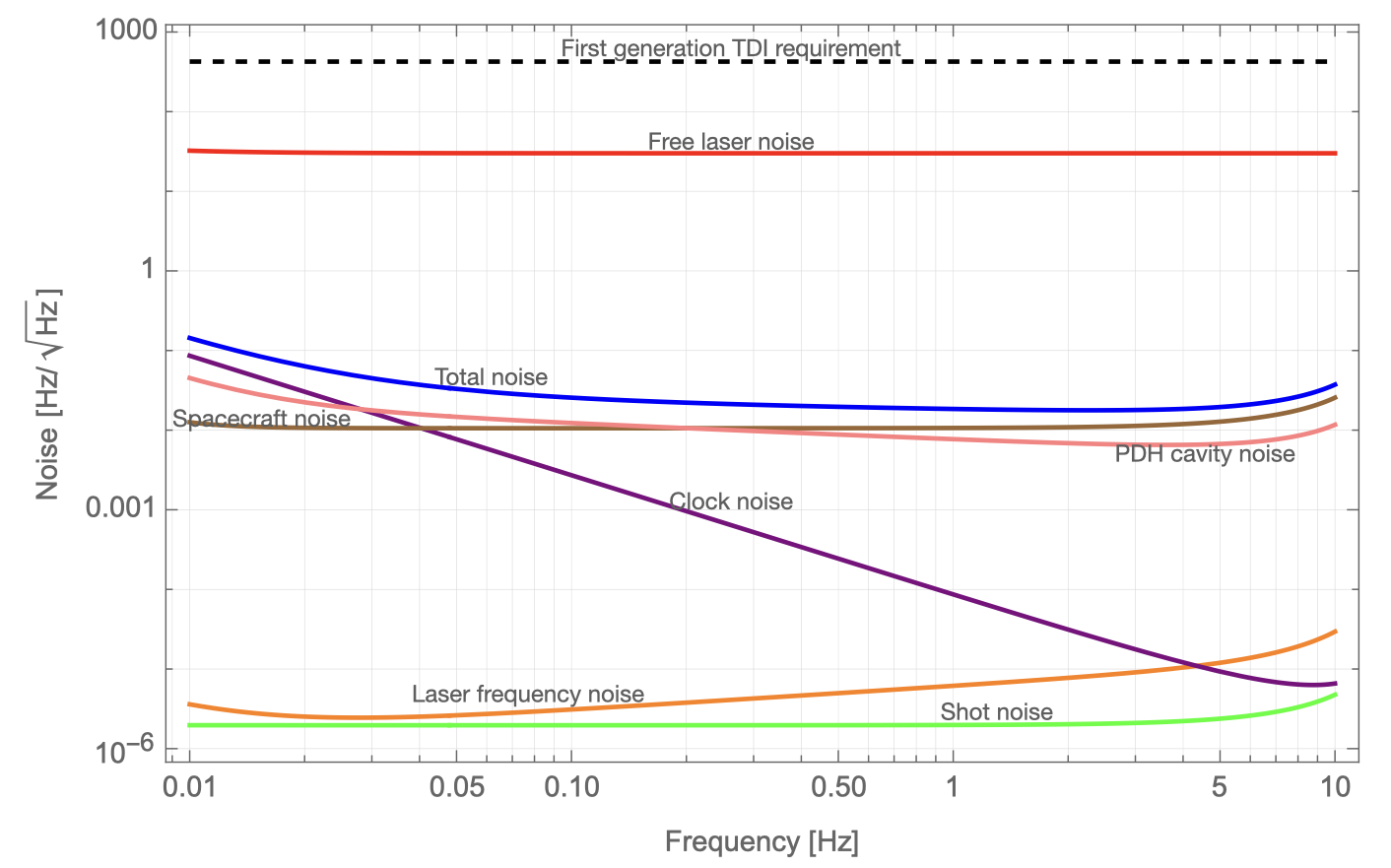}
        \caption{The noise budget of the double arm locking with PDH cavity for AMIGO}
        \label{fig:DoubleLockingAMIGOwPDH}
      \end{center}
    \end{figure}

\begin{figure}[!htb]
      \begin{center}
        \includegraphics[width=0.75\textwidth]{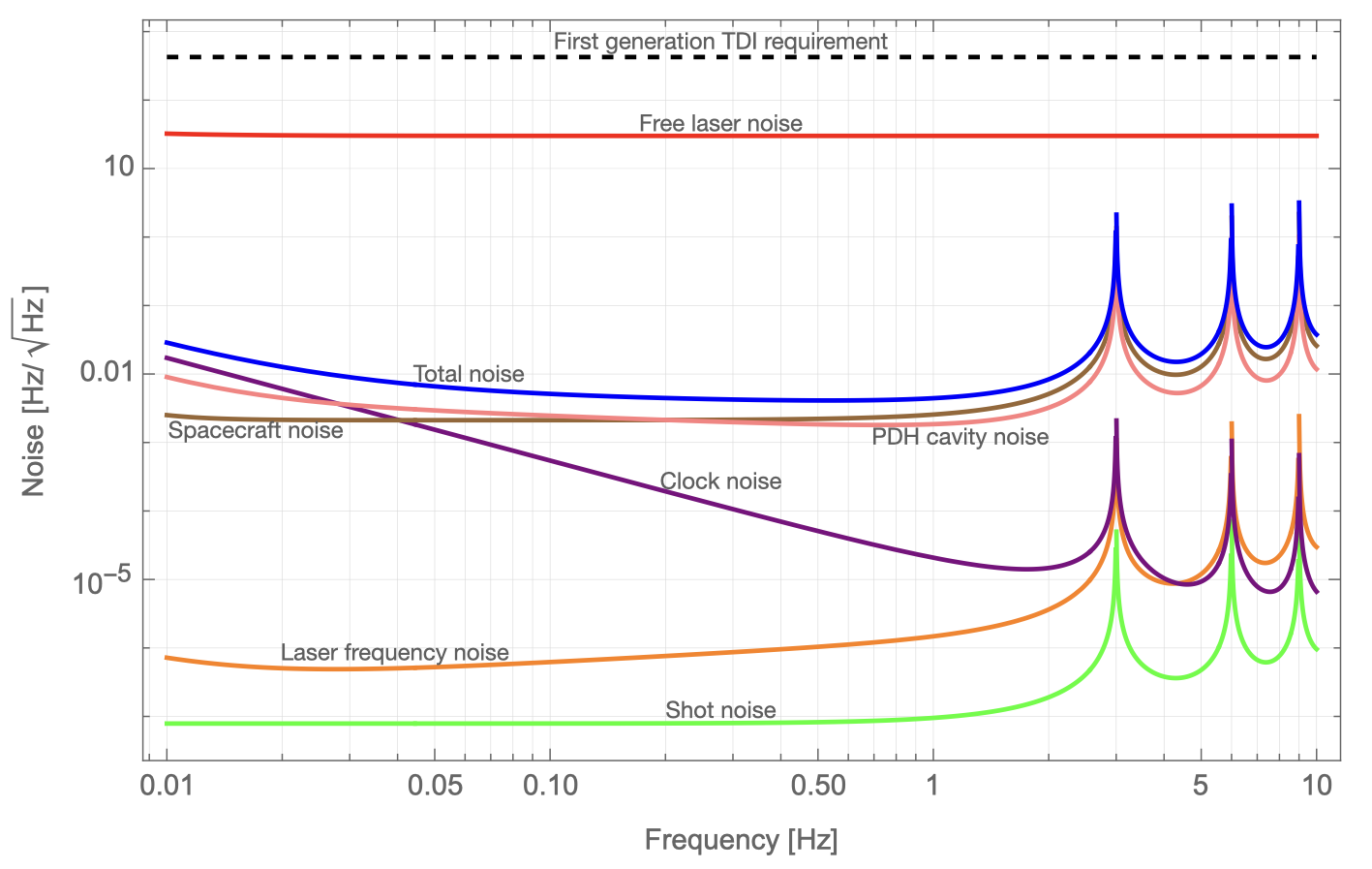}
        \caption{The noise budget of the double arm locking with PDH cavity for AMIGO-5}
        \label{fig:DoubleLockingAMIGO-5wPDH}
      \end{center}
    \end{figure}

The results after the common double arm locking enhanced by the PDH cavity are shown in Figs.~\ref{fig:DoubleLockingAMIGOwPDH} and \ref{fig:DoubleLockingAMIGO-5wPDH}. The convention of the lines is the same as the double arm locking case in the previous section. As before, we assume the average arm lengths for AMIGO and AMIGO-5 are 10,000 km and 50,000 km, respectively, and the difference between the two arms can be up to 1\%.

\newpage

With the common double arm locking enhanced by the PDH cavity,  the noise budgets for AMIGO and AMIGO-5 are plotted in Figs.~\ref{fig:DoubleLockingAMIGOwPDH} and \ref{fig:DoubleLockingAMIGO-5wPDH}.  From these figures,  we see that the total noise can be efficiently suppressed to meet the requirement of the first-generation TDI in the mid-frequency range.

\section{Doppler Frequency Pulling}\label{sec:FrequencyPulling}

\begin{figure}[!htb]
      \begin{center}
        \includegraphics[width=0.55\textwidth]{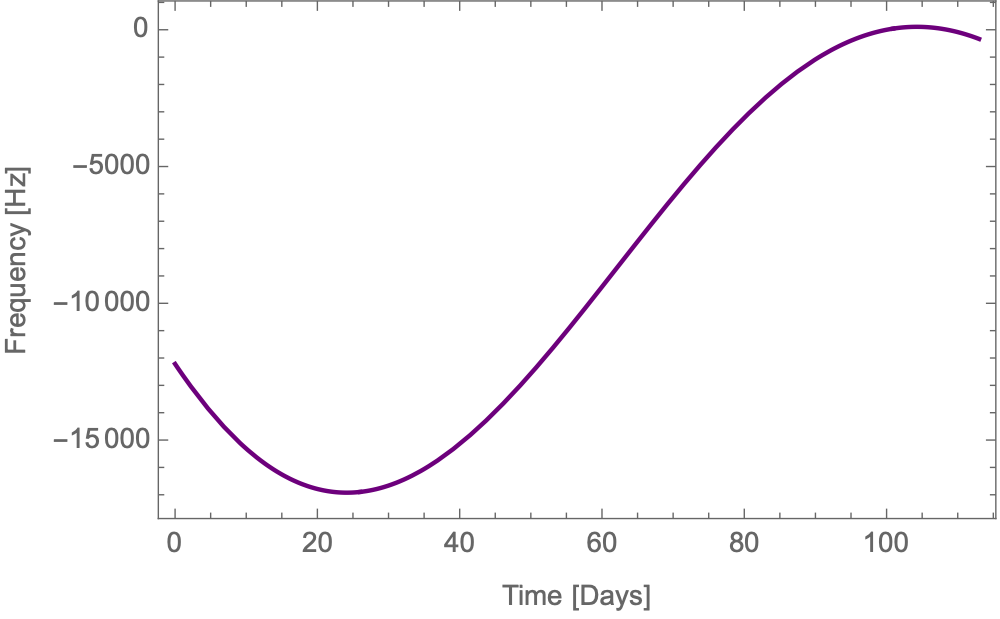}
        \caption{The early-time Doppler frequency noise model for AMIGO and AMIGO-5}
        \label{fig:DopplerNoise1}
      \end{center}
    \end{figure}

Suppose that the genuine Doppler frequency noise can be described by a sinusoidal model:
\be
  {\nu_{D,+}\left(t\right)=\nu}_1\ \sin{(\omega_1\ t+\phi_1)}+\nu_2\sin{(\omega_2\ t+\phi_2)}.
\ee
Before the arm locking is turned on, we can estimate the Doppler frequency noise using the following approximation:
\be
  \nu_{D;est\ }\left(t\right)=\nu_{0;+\ }+\gamma_{0;+}\ t+\int_{0}^{t}\int_{0}^{t^\prime} \alpha\left(t^{\prime\prime}\right)\ dt^{\prime\prime}\ dt^{\prime}\, ,
\ee
where
\be
  \alpha\left(t\right)=\alpha_1\ \sin{({\hat{\omega}}_1\ t+{\hat{\phi}}_1)}+\alpha_2\sin{({\hat{\omega}}_2\ t+{\hat{\phi}}_2)}\ .
\ee
For AMIGO and AMIGO-5, we show the Doppler frequency noise model and its estimate in Fig.~\ref{fig:DopplerNoise1} with the parameters in Tab.~\ref{tab:DopplerNoiseModel},  where we pick up a part of the early stage of Doppler velocity in Fig.~\ref{fig:Orbits}.

\begin{table}[htb!]
\centering
\begin{tabular}{|c|c|}
\hline
Parameter & Value\\
\hline
$\nu_1$ & $5.75807\times{10}^4$ \\
\hline
$\nu_2$ & $5.49284\times{10}^4$ \\
\hline
$\omega_1$ and $\hat{\omega}_1$ & $2\pi\times63.4\times{10}^{-9}$ rad/s \\
\hline
$\omega_2$ and $\hat{\omega}_2$ & $2\pi\times31.7\times{10}^{-9}$ rad/s \\
\hline
$\phi_1$ & $-1.62513$ rad \\
\hline
$\phi_2$ & $2.09707$ rad \\
\hline
$\nu_{0;+}$ & $-{10}^4$ \\
\hline
$\gamma_{0;+}$ & $-0.0067413$ \\
\hline
$\alpha_1$ & $9.13725\times{10}^{-9}$ \\
\hline
$\alpha_2$ & $-2.17909\times{10}^{-9}$ \\
\hline
${\hat{\phi}}_1$ & $1.51646$ rad \\
\hline
${\hat{\phi}}_2$ & $2.09707$ rad \\
\hline
\end{tabular}
\caption{The parameters in the Doppler frequency noise model \label{tab:DopplerNoiseModel}}
\end{table}

When we turn on the arm locking, it takes some time for the approximation to converge to the genuine Doppler frequency noise. To see this transient effect, we compute the following quantity in the time domain as an inverse Laplace transform
\be\label{eq:DopplerFrequencyPulling}
  \nu_C\left(t\right)=L^{-1}\ \left(\frac{\left[\nu_{D;+}\left(s\right)-\nu_{D;est\ }(s)\right]\ V(s)}{s}\right)\, ,
\ee
which is called the Doppler frequency pulling \cite{McKenzie:2009kt,  PhysRevD.90.062005,  Valliyakalayil:2021jxd}.  The arm locking transfer function $V(s)$ is given by
\be
  V(s)=-\frac{G_1(s)}{1+G_1\ P_++G_2\ P_{pdh}}\, ,
\ee
which depends on the arm length and other parameters of the arm locking. Here, we adopt the same choice of controllers $G_{1, 2}$ and $P_{pdh}$ as the combined double arm locking with PDH cavity discussed in the previous sections. The effect of the Doppler frequency pulling originates from the discrepancy between the genuine Doppler frequency noise and the noise modeling. However, this is also an effect in real observation that one has to consider because we need to know how fast the Doppler frequency noise model converges to the genuine Doppler noise in the presence of arm locking.

\begin{figure}[!htb]
      \begin{center}
        \includegraphics[width=0.55\textwidth]{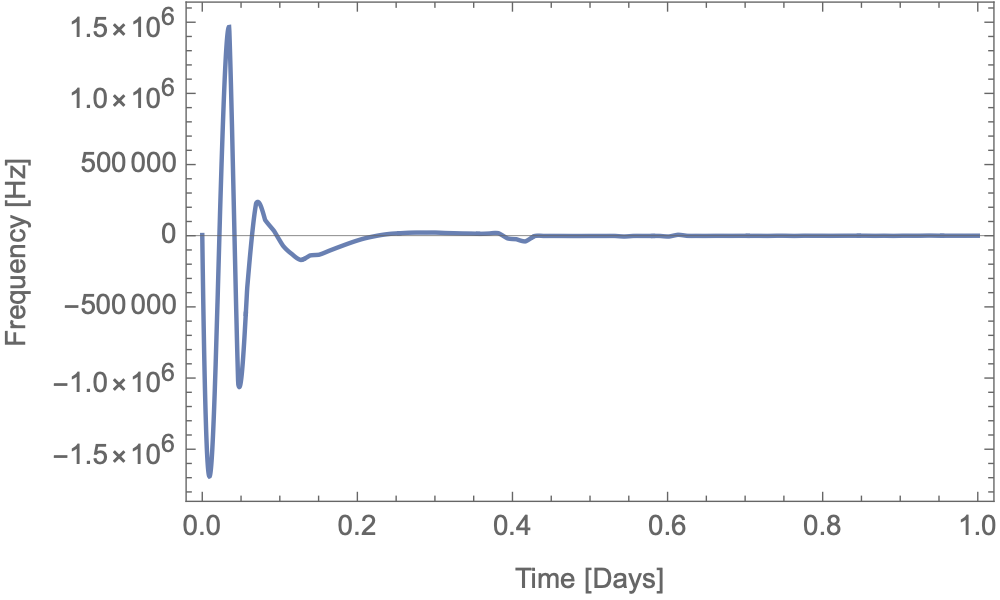}
        \caption{The result of the early-time Doppler frequency pulling for AMIGO}
        \label{fig:FrequencyPulling1AMIGO}
      \end{center}
    \end{figure}

\begin{figure}[!htb]
      \begin{center}
        \includegraphics[width=0.55\textwidth]{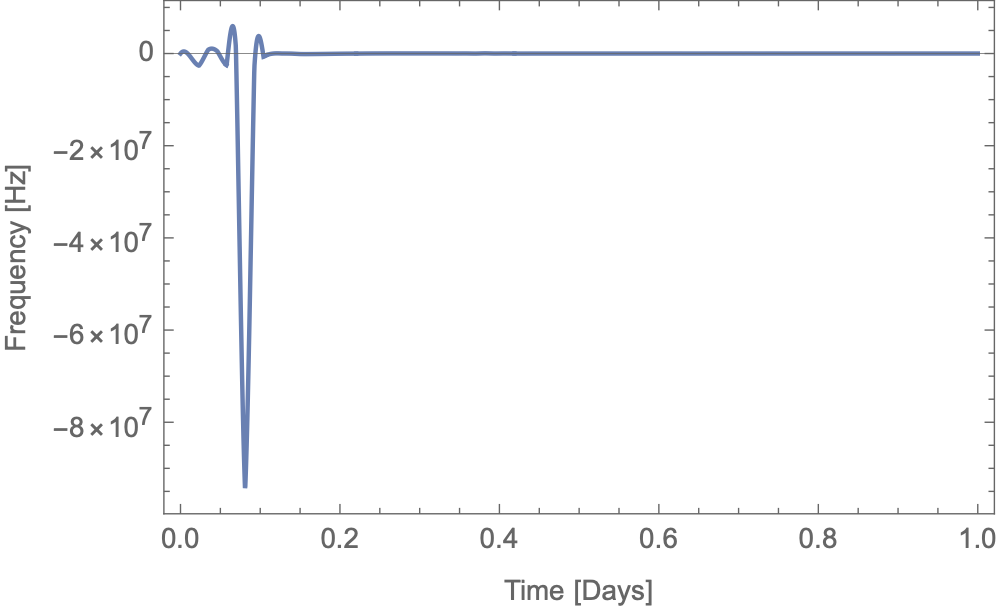}
        \caption{The result of the early-time Doppler frequency pulling for AMIGO-5}
        \label{fig:FrequencyPulling1AMIGO-5}
      \end{center}
    \end{figure}

\begin{figure}[!htb]
      \begin{center}
        \includegraphics[width=0.55\textwidth]{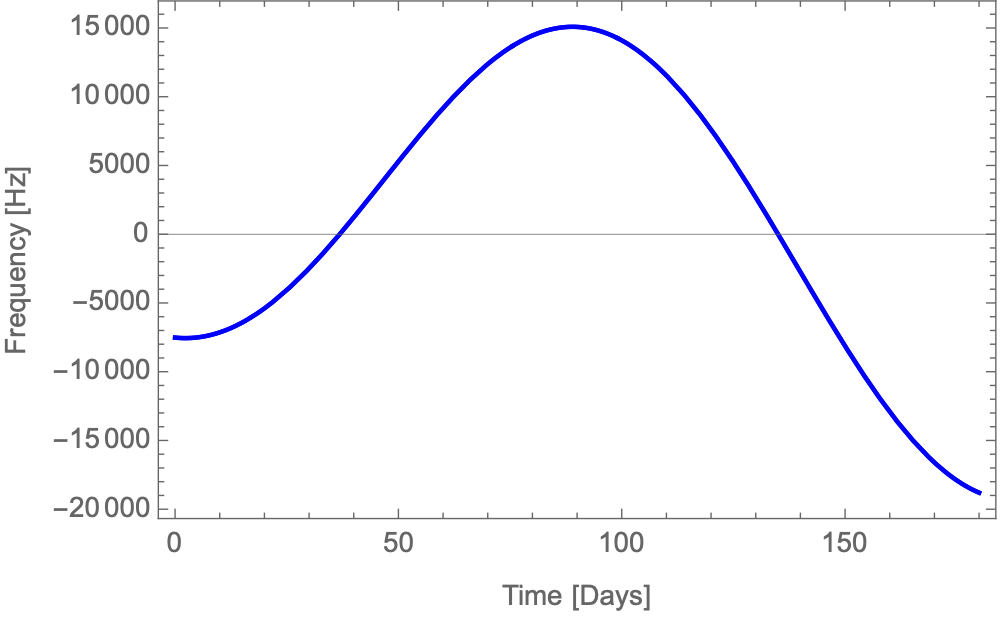}
        \caption{The intermediate-time Doppler frequency noise model for AMIGO and AMIGO-5}
        \label{fig:DopplerNoise2}
      \end{center}
    \end{figure}

\begin{figure}[!htb]
      \begin{center}
        \includegraphics[width=0.54\textwidth]{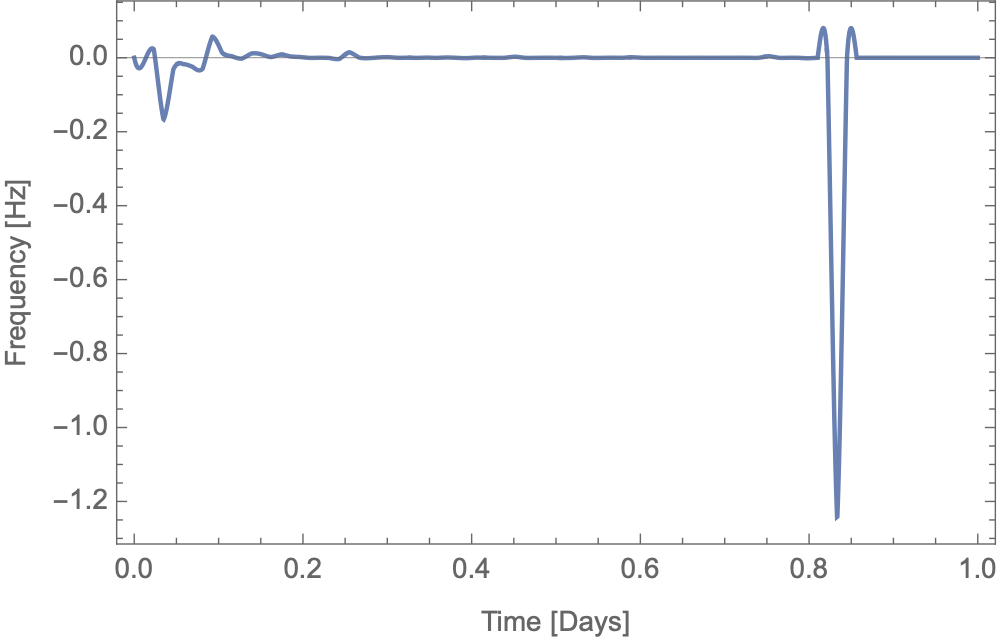}
        \caption{The result of the late-time Doppler frequency pulling for AMIGO-5}
        \label{fig:FrequencyPulling3AMIGO-5}
      \end{center}
    \end{figure}

For AMIGO and AMIGO-5, the results of the early-time Doppler frequency pulling are plotted in Figs.~\ref{fig:FrequencyPulling1AMIGO} and \ref{fig:FrequencyPulling1AMIGO-5}.  There is no fundamental difference between the early-time and the late-time Doppler frequency pulling.  In this paper,  we select several representative periods at different stages of the orbits for AMIGO and AMIGO-5 to demonstrate the universality of the Doppler frequency-pulling mechanism.  We can repeat the same steps and analyze the frequency pulling for later-time Doppler noises, and the results are shown in Figs.~\ref{fig:DopplerNoise2}$\, - \,$\ref{fig:FrequencyPulling3AMIGO-5}.

   \begin{figure}[!htb]
      \begin{center}
        \includegraphics[width=0.55\textwidth]{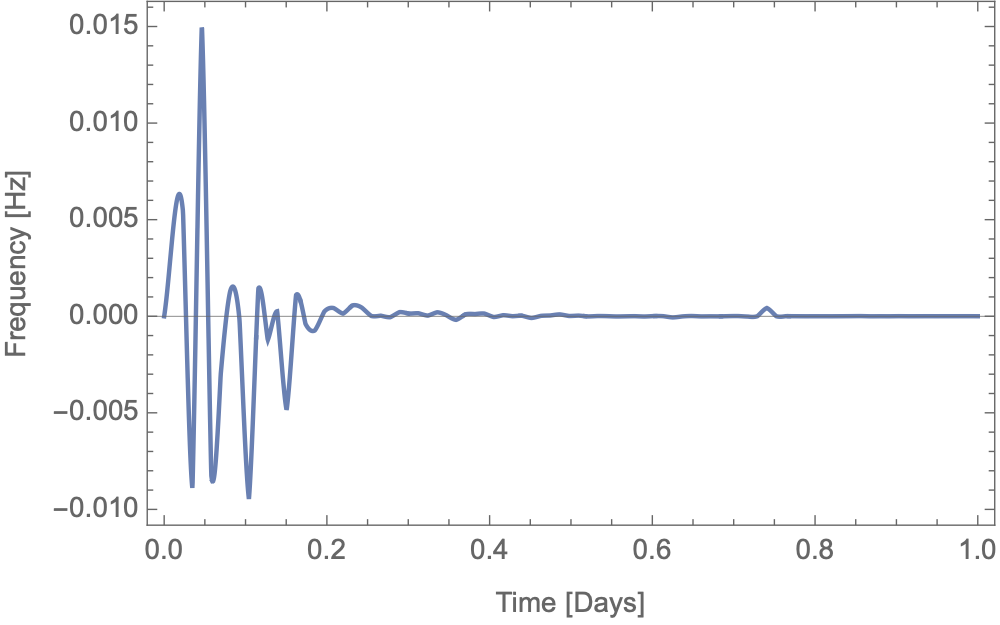}
        \caption{The result of the intermediate-time Doppler frequency pulling for AMIGO}
        \label{fig:FrequencyPulling2AMIGO}
      \end{center}
    \end{figure}

   \begin{figure}[!htb]
      \begin{center}
        \includegraphics[width=0.55\textwidth]{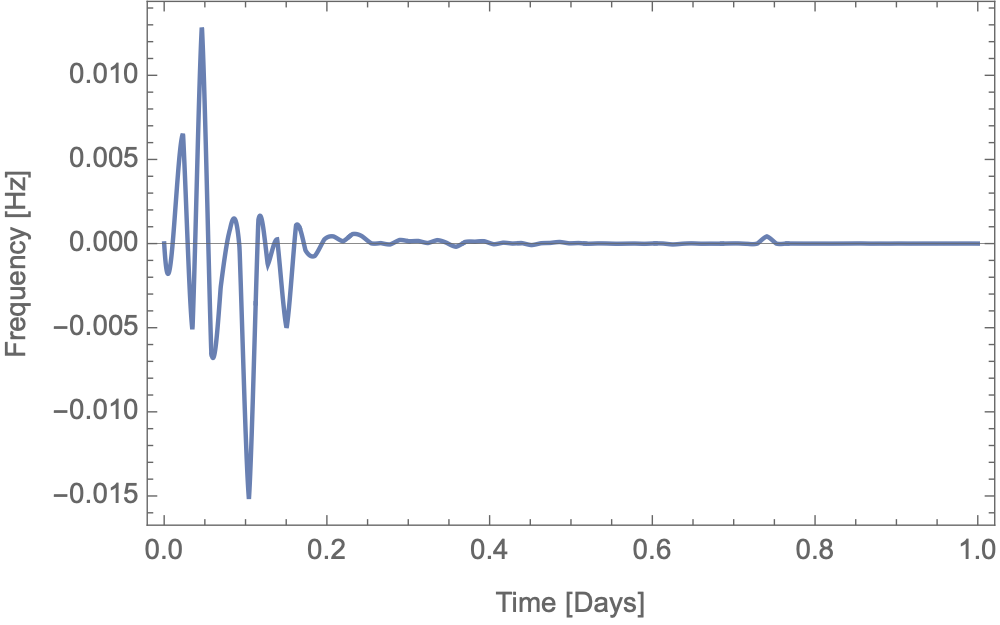}
        \caption{The result of the intermediate-time Doppler frequency pulling for AMIGO-5}
        \label{fig:FrequencyPulling2AMIGO-5}
      \end{center}
    \end{figure}

   \begin{figure}[!htb]
      \begin{center}
        \includegraphics[width=0.54\textwidth]{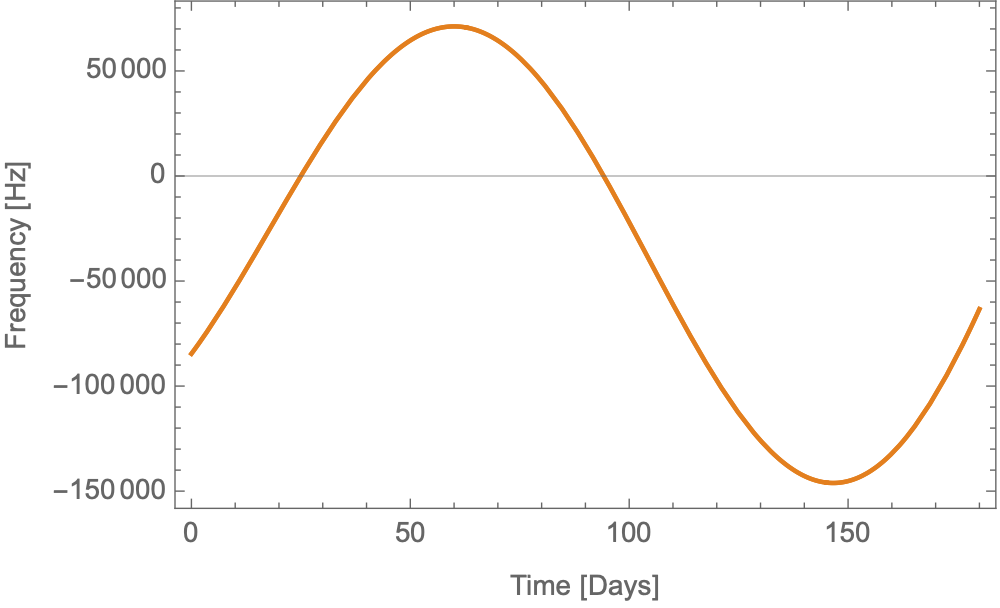}
        \caption{The late-time Doppler frequency noise model for AMIGO and AMIGO-5}
        \label{fig:DopplerNoise3}
      \end{center}
    \end{figure}

   \begin{figure}[!htb]
      \begin{center}
        \includegraphics[width=0.54\textwidth]{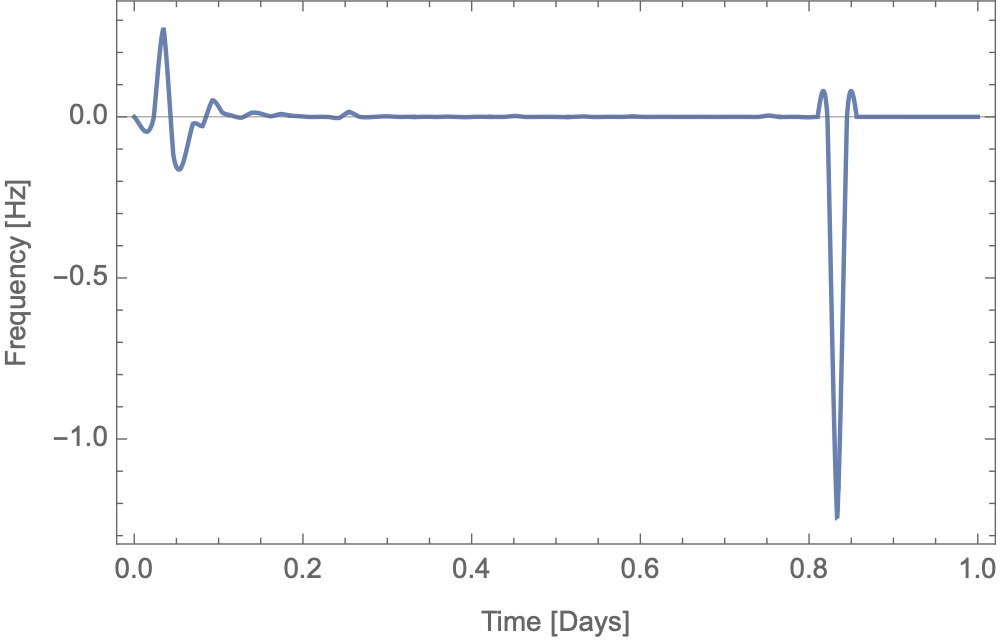}
        \caption{The result of the late-time Doppler frequency pulling for AMIGO}
        \label{fig:FrequencyPulling3AMIGO}
      \end{center}
    \end{figure}

\newpage
We see that the transient times for AMIGO and AMIGO-5 are both roughly $\sim$ 1.0 day, and the maximal magnitude ranges from $0.1$ Hz to $10^7$ Hz, depending on the stages of the Doppler velocities.  The jumps in the figures of Doppler frequency pulling (Figs.~\ref{fig:FrequencyPulling1AMIGO},  \ref{fig:FrequencyPulling1AMIGO-5},  \ref{fig:FrequencyPulling2AMIGO},  \ref{fig:FrequencyPulling2AMIGO-5},  \ref{fig:FrequencyPulling3AMIGO},  \ref{fig:FrequencyPulling3AMIGO-5}) indicate the largest mismatch between the genuine and the approximated Doppler frequency noises, when the arm locking is turned on. In fact, we computed the Doppler frequency pulling numerically in the time domain using \eqref{eq:DopplerFrequencyPulling} for much longer time ($\sim 100$ days), but the maximal discrepancy always happens within 1.0 day. Hence, we conclude that after this transient time ($\sim 1.0$ day), the pulling will reach a steady state limited to $\pm$ 0.001 Hz. Hence, the arm locking technique can efficiently reduce the discrepancy between the genuine Doppler frequency noise and the noise model within a relatively short period of time, and the Doppler frequency noise model is trustworthy after 1.0 day.

\section{Discussion}\label{sec:Discussion}

In this paper, we study the arm locking technique applied to mid-frequency gravitational wave detection.  More specifically,  we consider the Astrodynamical Middle-frequency Interferometric Gravitational wave Observatory (AMIGO) as a prototype to demonstrate the efficiency of the arm locking technique in laser noise suppression.

For AMIGO, the first-generation TDI requirement is already satisfied by more than two orders of magnitude. For the original Michelson topology (the zeroth-generation) TDIs, there is still a gap of around six orders of magnitude.  We look for arm locking to come to the rescue or partial rescue.  To find the efficiency of the arm locking technique, we have analyzed the suppression of various noises in the frequency domain. In both cases, with various concrete configurations, we have seen that the arm locking can suppress the laser frequency noise by about three orders of magnitudes. This way, the gap can be bridged by three orders of magnitude. If the thrust-acceleration method is used for minimizing the inequality of the arm lengths, the requirement on the thrust and inertia sensor/accelerometer is relaxed quite a bit \cite{Ni:2019nau}.  We have considered the frequency pulling caused by the Doppler noise. By optimizing the parameters, we can reduce the transient time to $\sim$ 0.5 days for AMIGO and $\sim$ 1.0 day for AMIGO-5, demonstrating the practical feasibility of using arm locking to reduce laser-frequency noise for mid-frequency gravitational wave detection.

Although the arm locking technique has been studied in the previous literature (e.g.  for LISA),  for the first time we apply it to the mid-frequency range,  which has not been explored before.  Meanwhile,  we gain some new insights of this technique.  For example,  we found that the parameters of the controllers can be optimized to minimize the Doppler frequency pulling time.

\section*{Acknowledgments}

We thank the referees for their critical readings of the manuscript and helpful comments. We also thank Yong Tang for helpful discussions. This work is supported by the National Key Research and Development Program of China (Grant No.~2021YFC2201901) and in part by the NSFC under grant No.~12147103.  J.N.  would like to thank the International Centre for Theoretical Physics (ICTP) and the Korea Institute for Advanced Study (KIAS) for the warm hospitality during the final stage of this work.

%%%%%%%%%%%%%%%%%%%%%%%%%%%%%%%%%%%%%%%%%%%%%%%%%%
%%%%%%%%%%%%%%%%%%%%%%%%%%%%%%%%%%%%%%%%%%%%%%%%%%
%\bibliographystyle{plain}
\bibliographystyle{utphys}
\bibliography{ArmLocking}
\end{document}